\documentstyle[prb,aps,multicol,epsf]{revtex}
\tighten
\begin{document}
\draft
\title{
Universality of the Kondo Effect in a Quantum Dot out of Equilibrium}
\author{A. Kaminski$^1$, Yu.V. Nazarov$^2$, L.I. Glazman$^1$}
\address{$^1$Theoretical Physics Institute, University
of Minnesota, Minneapolis, MN 55455, USA\\
$^2$Delft University of Technology, 2600 GA Delft, The
Netherlands}
\maketitle
\begin{abstract}
  We study the Kondo effect in a quantum dot driven out of equilibrium
  by an external ac field.  The Kondo effect can be probed by
  measuring the dc current induced by an auxiliary dc bias $V_{\rm dc}$
  applied across the dot.  In the absence of ac perturbation, the
  corresponding differential conductance $G(V_{\rm dc})$ is known to
  exhibit a sharp peak at $V_{\rm dc}=0$, which is the manifestation
  of the Kondo effect.  In the equilibrium, there exists only one
  energy scale, the Kondo temperature $T_K$, which controls all the
  low-energy physics of the system; $G$ is some universal function of
  $eV_{\rm dc}/T_K$.  We demonstrate that the dot out of equilibrium
  is also characterized by a universal behavior: conductance $G$
  depends on the ac field only through two dimensionless parameters,
  which are the frequency $\omega$ and the amplitude of the ac
  perturbation, both divided by $T_K$. We find analytically the large-
  and small-frequency asymptotes of the universal dependence of $G$ on
  these parameters. The obtained results allow us to predict the
  behavior of the conductance in the crossover regime $\hbar\omega\sim
  T_K$.
\end{abstract}

\begin{multicols}{2}
\section{Introduction}

The Kondo effect results from exchange interaction of itinerant
electrons with a localized spin state. This interaction leads to local
spin polarization of the electron gas. The polarization becomes
significant only at low temperatures, due to the existence of
collective states\cite{Hewson} with small excitation energies.
Simultaneously with the modification of the spin susceptibility, the
scattering properties are modified significantly from those of the
original localized spin state.  The modification is especially
striking in the case of antiferromagnetic exchange interaction, when
the ``spin cloud'' formed out of free electron gas tends to screen the
magnetic moment of the localized state.  In this case, the scattering
cross-section grows as the temperature is being lowered, and reaches
the unitary limit at $T\to 0$. This phenomenon is responsible for the
non-monotonic temperature behavior of the resistivity of metals with
magnetic impurities at low temperatures, which was the first
experimentally observed manifestation of the Kondo
resonance.\cite{Hewson} However, the system of magnetic impurities
embedded in a metal sample does not offer much control over the
parameters even at the stage of the fabrication of the sample, not
speaking of the dynamic variation of the parameters in the course of
an experiment.

Another class of systems, whose transport properties can also be
affected by the Kondo effect, but which offer a much higher level of
control over the system parameters, is provided by quantum dots.  A
quantum dot in a semiconductor planar heterostructure is a confined
few-electron system contacted by sheets of two-dimensional
electron gas (leads).  If the total number of electrons on the dot is
odd, then the dot is similar to a magnetic impurity.  Junctions
between the dot and the leads produce an overlap of the states in the
dot and in the 2D leads.  This overlap leads to the exchange
interaction between the spin of the dot and spins of the itinerant 2D
electrons. At sufficiently low temperatures, the ``spin cloud'' is
formed of the electrons in the leads. Like in bulk metals, the
scattering off the resulting many-body state is being enhanced as the
temperature is being lowered, and reaches unitary limit at $T\to 0$.
Different are only the manifestations of the Kondo resonance in bulk
metals and in the quantum dot systems. In bulk metals, the enhancement
of scattering by the Kondo resonance increases the resistivity. In
quantum dot systems, on the contrary, the scattering facilitates
transport through the dot. Kondo effect results in a specific
temperature dependence of the linear conductance across the dot.  If
temperature is lowered, the conductance first drops due to the Coulomb
blockade phenomenon,\cite{Kouwenhovenreview} and then increases again
due to the Kondo effect.\cite{GlazmanRaikh88} At $T\to 0$, the
conductance $G$ reaches its maximum, which corresponds to the unitary
limit of tunneling.  Recently, the Kondo effect in a quantum dot was
observed experimentally.\cite{GoldhaberEtal98,CronenwettEtal98,Weis}

The quantum dot devices are highly controllable, and can be operated
in regimes inaccessible in the conventional magnetic impurity systems,
that were used previously for studying the Kondo effect. Irradiation
of a quantum dot with an ac field offers a new, clever way of
affecting its dynamics, which enables one to study the Kondo anomaly
in essentially non-equilibrium conditions. The ac field can be applied
to the gate, thus modulating the dot's potential with respect to the
leads; alternatively, one may apply ac bias to the leads. In any case,
driving the system out of equilibrium affects the dc conductance
discussed above. Measuring the dc $I-V$ characteristics, one can
investigate the effects of the irradiation on the Kondo anomaly.

A generic theoretical description of a quantum dot operates with a
significant number of parameters and energy scales describing the
system.  Nevertheless, in the case of no ac field, it turns out that
the low-energy properties of the quantum dot system which are related
to the Kondo effect are controlled by only one relevant energy scale,
which is the Kondo temperature $T_K$. The Kondo temperature, in turn,
depends on the microscopic parameters of the system, {\em e.g.}, on
the gate voltage and conductances of the dot-lead junctions. Such
universality allows for easier understanding and description of the
problem.

The ac field introduces new parameters to the problem, thus apparently
breaking the universal description, which is valid in the static case.
This re-emerging abundance of parameters makes it difficult to build a
consistent description of the effects of the ac field on the Kondo
conductance.  The theoretical work performed up to the date, was
concentrated on some specific regimes. Goldin and
Avishai\cite{GoldinAvishai98} considered the case of very strong ac
bias with the help of the third-order perturbation theory in the
dot-lead coupling. Nordlander {\em et al.}\cite{NordlanderEtal98}
analyzed the effects of ac field of sufficiently high frequency, which
ionizes the dot. They have conjectured that even at the temperature of
the thermal bath $T=0$, the finite rate of the dot ionization results
in a finite effective temperature ``seen'' by the Kondo state. This
way, irradiation provides the cut-off for the Kondo singularity and
reduces the conductance. Later, we have
demonstrated\cite{KaminskiEtal99} that even in the absence of the dot
ionization, irradiation is able to flip the spin of the dot, thus
bringing decoherence into the Kondo state and diminishing the Kondo
effect. In addition to analytical methods, a number of numerical
approaches have been used\cite{NordlanderEtal98,LopezEtal98} to study
the conductance of a Kondo system out of equilibrium at certain sets
of values of the bare parameters of the system. Because of the large
number of the parameters involved, the results of such calculations
are hard to analyze. At any rate, such a limited consideration could
not reveal the universalities of the problem. In our view, it also
does not provide an insight into the regimes which do not allow for a
perturbative treatment.

In our earlier paper\cite{KaminskiEtal99} we have learned how to apply
the Renormalization Group (RG) technique for a Kondo system out of
equilibrium. It allowed us to sum up the infinite series of the
perturbation theory in the dot lead coupling. This treatment, valid at
relatively high decoherence rates, yielded the expression for the
conductance as a function of only one parameter, which is the ratio of
the decoherence rate $\hbar/\tau$ to the Kondo temperature.  Thus we
established that the Kondo temperature $T_K$ remains a meaningful
parameter of the theory, even if the ac field strongly suppresses the
manifestations of the Kondo effect. We do not see a way, however, of
measuring the decoherence rate directly. Moreover, the definition of
this quantity in Ref.~\onlinecite{KaminskiEtal99} makes sense only at
a sufficiently high frequency of the ac field, $\hbar\omega\gg T_K$.
Therefore, the dimensionless ratio $\hbar/\tau T_K$ can not be the
only parameter describing the effect of ac field on the Kondo system.

In this paper, we find the correct dimensionless variables that
characterize the amplitude and frequency of the ac field perturbing
the Kondo system. If the ac field biases the dot, then the proper
variables are $eV_{\rm ac}/T_K$ and $\hbar\omega/T_K$, where $V_{\rm
  ac}$ is the amplitude of the ac bias. In the case of ac modulation
of the gate voltage, the perturbation introduced by the ac field is
characterized by the corresponding dimensionless variation of the
Kondo temperature $\delta T_K/T_K$, which in principle can be
independently measured. In terms of the proper pair of
variables, the behavior of the Kondo conductance is universal.  We
find analytically the asymptotes of the universal dependence by
further developing the RG treatment (valid in the case of strong
suppression of the Kondo effect), and by generalizing the
Nozi\`{e}res' Fermi-liquid theory\cite{Nozieres} onto the
non-equilibrium case (which adequately describes the limit of weak
perturbation by a low-frequency field).

Within this picture, we are able to describe in a consistent way the
effect of irradiation in a wide range of frequencies of the ac
perturbation - from zero to the dot ionization threshold; this
includes the most interesting for the current experiments\cite{exp} region
$\hbar\omega\sim T_K$. The remarkable feature of the presented
description is that the Kondo temperature remains the only relevant
energy scale, despite the essentially non-equilibrium character of the
problem.

The outline of our paper is as follows. In Sec.~\ref{sec:Hams} we
introduce the description of the system by means of the time-dependent
Kondo Hamiltonian. Then we derive the expression for the Kondo
conductance of the dot in the absence of ac field. The purpose of this
derivation is to present the formalism which later would be suitable
to describe the non-equilibrium states produces by the irradiation.

In Sections~\ref{sec:weak}--\ref{sec:sum1} we consider the effect of
ac modulation of the gate voltage on the Kondo conductance.
At higher frequencies, $\hbar\omega/T_K\gtrsim 1$,
the decoherence of the Kondo many-body state is the principal channel
via which the ac field influences the Kondo anomaly. This frequency
domain is considered in Sections~\ref{sec:weak}--\ref{sec:crossover}.
Depending on the strength of modulation, $\delta T_K/T_K$, the
suppression of the Kondo conductance is significant
(Section~\ref{sec:weak}), or relatively weak
(Sections~\ref{sec:crossover}).  In Sec.~\ref{sec:strong} we consider
the limit of very small frequencies of the ac field. The decoherence
probability in these conditions is exponentially small.  However, the
adiabatic evolution of the collective state, caused by the ac modulation,
affects its scattering properties.  It results in small deviation of
the conductance from the unitary limit. The results of Sections
\ref{sec:weak}--\ref{sec:strong} are summarized in
Sec.~\ref{sec:sum1}.

In Sec.~\ref{sec:weakacbias} we consider the effect of ac bias on the
Kondo conductance. 
It turns
out that a strong suppression of the Kondo effect is possible only if
$eV_{\rm ac}/T_K\gg 1$. Upon the increase of the frequency, the
suppression diminishes. This frequency dependence is opposite to the
one in the case of gate voltage modulation.

In Sec.~\ref{sec:sat} we consider the satellite peaks which can be
created by irradiation in the $I$--$V$ characteristic of a quantum
dot.

Finally in Sec.~\ref{sec:exp} we compare
experimental results\cite{exp} with our theory.

\section{The Kondo effect in a quantum dot}
\label{sec:Hams}

\subsection{The model}
\label{sec:AnHam}

The system we study is a quantum dot attached to two leads by
high-resistance junctions, so that the charge of the dot is nearly
quantized. The Kondo effect emerges in a quantum dot occupied by an
odd number of electrons at temperatures below the mean level spacing
in the dot.  Under such conditions, the topmost occupied level is
special, since it is filled by only one electron. It is this level
which produces the Kondo effect. The other levels, occupied by two or
no electrons are unimportant in our discussion (similarly to the inner
shells of a magnetic impurity in the conventional Kondo effect).
Therefore the model of the dot attached to two leads can be truncated
to the Anderson single-level impurity model, 
\begin{mathletters}
\label{Hamiltonian}
\begin{eqnarray}
\hat{H}&=&\!\sum_{k,\sigma,\alpha}(\xi_k+eV_\alpha)
c^{\dagger}_{k\sigma\alpha}c^{\phantom{\dagger}}_{k\sigma\alpha}
+
\!\sum_{k,\sigma,\alpha}(v_\alpha
c^{\dagger}_{k\sigma\alpha}d^{\phantom{\dagger}}_{\sigma}+
{\rm H.c.}
)\nonumber\\
&+&\sum_\sigma (-E_d+eV_{\rm dot}\cos\omega t) d^{\dagger}_{\sigma}
d^{\phantom{\dagger}}_{\sigma}+
Ud^{\dagger}_{\uparrow}d^{\phantom{\dagger}}_{\uparrow}
d^{\dagger}_{\downarrow}d^{\phantom{\dagger}}_{\downarrow}\;;
\label{H}\\
&&V_{L,R}=\pm\textstyle\frac{1}{2}\left[V_{\rm dc}+
V_{ac}\cos(\omega't+\phi_0)\right]\;,\nonumber\\
&&\Gamma_\alpha\equiv 2\pi \nu  |v_\alpha|^2\;.
\label{Gamma}
\end{eqnarray}
\end{mathletters}
Here the first two terms correspond to non-interacting electrons in the two
leads ($\alpha=L,R$), and tunneling of free electrons between the dot
and the leads, respectively. The dot is described by the third and
fourth terms of the Hamiltonian, $E_d$ and $U-E_d$ are the ionization
and the electron addition energy, respectively. The tunneling matrix
elements $v_\alpha$ are related to the widths $\Gamma_\alpha$ by
Eq.~(\ref{Gamma}), where $\nu $ is the density of states in a lead.
The ac field can be applied to the gate, which is coupled to the dot
capacitively, and thus can modulate the energy of the electron
localized in the dot with the amplitude $eV_{\rm dot}$. 
 We assume that
the leads can be either only dc-biased, or an additional ac bias can
be applied.

\subsection{Time-dependent Schrieffer-Wolff transformation}
\label{sec:Kondoham}

In the present paper we consider the dot in the Kondo regime,
$U-E_d,\, E_d\gg \Gamma_{L,R}$. Under such conditions, the number of
electrons on the dot is a well-defined quantity. In the limit of
infinitely small tunneling, the ground state of the system described
by the Hamiltonian (\ref{Hamiltonian}) is doubly degenerate due to the
spin of the (single) electron which occupies the level $d$.  The
states with two or no electrons on the dot are higher in energy by
$U-E_d$ or $E_d$ respectively, and are not important for the
low-energy dynamics of the system.  In our paper, we 
study the irradiation effects when the applied fields do not
drive the dot out of the Kondo regime,
\begin{equation}
eV_{\rm dc},\;eV_{\rm dot},\; eV_{\rm ac} < E_d,\;U-E_d\;.
\end{equation}
Therefore, the excited states with two or no electrons on the dot are
to be projected out. This can be
achieved by the Schrieffer-Wolff
transformation,\cite{SchriefferWolff66} modified to account for the
time dependence of the parameters of the Hamiltonian
(\ref{Hamiltonian}).  In the present subsection we perform this
transformation, which finally yields the description of the quantum
dot system by means of the Kondo Hamiltonian with time-dependent
parameters.

First we move all dependence of the Hamiltonian on the applied
voltages $V_{\rm dot}$, $V_{\rm dc}$, $V_{\rm ac}$ to the off-diagonal
terms. It is achieved by the unitary transformation
\begin{eqnarray}
{\sf U}&=&\exp\left\{-\frac{ie}{\hbar}\int^t\! dt'
\left[ \sum_{k,\sigma,\alpha}V_\alpha(t')
c^{\dagger}_{k\sigma\alpha}c^{\phantom{\dagger}}_{k\sigma\alpha}
\right. \right.
\nonumber\\
&&\qquad\qquad\qquad\left.\left.
\vphantom{\int^t\sum_{k,\sigma,\alpha}}
+V_{\rm dot}(t')
d^{\dagger}_{\sigma}d^{\phantom{\dagger}}_{\sigma}\right]
\right\}\;.
\label{U1}
\end{eqnarray}
After this transformation, the Hamiltonian has the form
\begin{eqnarray}
\hat{H}'&=&{\sf U}\hat{H}{\sf U}^\dagger-
i\hbar\frac{\partial{\sf U}}{\partial t}{\sf
U}^\dagger\nonumber\\
&=&\sum_{k,\sigma,\alpha}\xi^{\phantom{\dagger}}_k
c^{\dagger}_{k\sigma\alpha}c^{\phantom{\dagger}}_{k\sigma\alpha}
+\sum_\sigma (-E_d) d^{\dagger}_{\sigma}
d^{\phantom{\dagger}}_{\sigma}+
Ud^{\dagger}_{\uparrow}d^{\phantom{\dagger}}_{\uparrow}
d^{\dagger}_{\downarrow}d^{\phantom{\dagger}}_{\downarrow}
\nonumber\\
&\phantom{+}&
+
\sum_{k,\sigma,\alpha}\left[\tilde{v}^{\phantom{\dagger}}_\alpha(t)
c^{\dagger}_{k\sigma\alpha}d^{\phantom{\dagger}}_{\sigma}+
{\rm H.c.}\right]
\;,
\label{Hprim}
\end{eqnarray}
with
\begin{equation}
  \label{tildev}
  v_\alpha(t)\equiv v_\alpha
\exp\left\{-\frac{ie}{\hbar} \int^t\!
dt'\left[V_\alpha(t')-V_{\rm dot}(t')\right]\right\}\;.
\end{equation}

Now we can make the time-dependent Schrieffer-Wolff transformation,
which is defined by the unitary operator 
\begin{equation}
  \label{sfW}
{\sf W}=\exp({\sf V})  
\end{equation}
with
\begin{eqnarray}
{\sf V}=\sum_{k,\sigma,\alpha}\left\{\left[
w_{k\alpha}^{(1)}(t)(1-n_{-\sigma})+w_{k\alpha}^{(2)}(t)n_{-\sigma}
\right]
d^\dagger_\sigma c^{\phantom{\dagger}}_{k\sigma\alpha}
\right.\nonumber\\
\left.\vphantom{a_{\alpha}^{(1)}}
-{\rm H.c.}\right\}\;.
\label{sfV}
\end{eqnarray}
The functions $w_{k\alpha}^{(j)}(t)$ are to be found from the condition
\begin{equation}
0=\hat{H}_v+[{\sf V},\hat{H}_1]-i\hbar\frac{\partial {\sf V}}{\partial
t}\;,\label{accondition}
\end{equation}
where $\hat{H}_v$ is the part of the Hamiltonian (\ref{Hprim})
responsible for mixing of electron states in the leads and in the dot
[the last term of Eq.~(\ref{Hprim})], and $\hat{H}_1$ describes the
uncoupled dot and leads [the first three terms in Eq.~(\ref{Hprim})].
The condition (\ref{accondition}) ensures that the resulting
Hamiltonian ${\sf W} \hat{H}' {\sf W}^\dagger$ has no
linear-in-$v_\alpha$ terms, which account for the variations of the
number of electrons in the dot.

The only difference of the transformation (\ref{sfW})--(\ref{sfV})
from the conventional Schrieffer-Wolff
transformation\cite{SchriefferWolff66} is the time dependence of
$w_{k\alpha}^{(j)}$. For the static Anderson Hamiltonian, these
factors are constant; in our case they are functions of time because
of the time variations of the Hamiltonian (\ref{Hamiltonian}).
Solving Eq.~(\ref{accondition}) for $w_{k\alpha}^{(1)}(t)$, we obtain
\begin{equation}
  \label{akalpha}
  w_{k\alpha}^{(1)}(t)=\left[-i\int^t dt' e^{i(\xi_k-E_d)t'/\hbar} 
v_\alpha(t')\right] e^{-i(\xi_k-E_d)t/\hbar}\;.
\end{equation}
When the applied ac fields are slow enough, $\hbar\omega/E_d,\
\hbar\omega'/E_d \ll 1$, one can solve Eq.~(\ref{accondition}) in the
adiabatic approximation, neglecting the third term in it. This approach
yields a simplified expression for $w_{k\alpha}^{(1)}(t)$:
\begin{equation}
  \label{akalphaadiab}
  w_{k\alpha}^{(1)}(t)\approx
\frac{v_\alpha(t)}{E_d-eV_\alpha(t)+eV_{\rm dot}(t)}\;.
\end{equation}
Here we have also neglected the single-electron energies $\xi_k$
because the Kondo effect is produced by the states close to the Fermi
level, whose energy is small in comparison to $E_d$.  The formulas for
$w_{k\alpha}^{(2)}(t)$ are analogous to
Eqs.~(\ref{akalpha})--(\ref{akalphaadiab}), only $E_d$ must be
replaced by $E_d-U$.

Applying transformation (\ref{sfW})--(\ref{sfV}) to
the Hamiltonian (\ref{Hprim}), we come to the Kondo Hamiltonian
\begin{mathletters}
\label{HK}
\begin{eqnarray}
\hat{H}_K&=&\hat{H}_0+\hat{H}_{\cal J}\;,\quad
\hat{H}_0=
\sum_{k,\sigma,\alpha}\xi_k^{\phantom{\dagger}}
c^{\dagger}_{k\sigma\alpha}
c^{\phantom{\dagger}}_{k\sigma\alpha}\;,
\label{HK0}\\
\hat{H}_{\cal J}&=&\!\sum_{\parbox{0.3in}{\scriptsize
$k,\sigma,\alpha$\\
$k'\!,\sigma'\!\!,\alpha'$}}\!
{\cal J}_{\alpha\alpha'}(t)
\left(\textstyle\frac{1}{4}\delta_{\sigma\sigma'}+
\hat{S}_l s^l_{\sigma\sigma'}\right)
c^{\dagger}_{k\sigma\alpha}
c^{\phantom{\dagger}}_{k'\sigma'\alpha'}\;,
\label{HKt}
\end{eqnarray}
\end{mathletters}
where ${\hat{\bf s}}$ and ${\hat{\bf S}}$ are the spin operators of
the electrons in the leads and of the electron on the isolated level,
respectively; we assume summation over the repeating indices
$l=x,y,z$.  In the framework of the Hamiltonian (\ref{HK}), the state
of the dot is fully described by its spin. The terms of the Anderson
Hamiltonian (\ref{Hamiltonian}) that are responsible for the electron
tunneling to/from the dot, and for the Coulomb interaction of
electrons in the dot have been transformed to the term $\hat{H}_{\cal
  J}$ of the Kondo Hamiltonian (\ref{HK}). This term represents
exchange interaction between the spin of the dot and the electrons in
the leads.  The Hamiltonian (\ref{HK}) operates within the band
$-E_d<\xi_k<U-E_d$, see Ref.~\onlinecite{Haldane}. The coupling
parameters ${\cal J}_{\alpha\alpha'}(t)$ are given by
\begin{equation}
  \label{J}
{\cal J}_{\alpha\alpha'}(t)=
\tilde{v}_\alpha(t)\left[-w_{k'\alpha'}^{(1)}(t)+w_{k'\alpha'}^{(2)}(t)\right]^*+
\mbox{C.c.}
\end{equation}
The applied bias is accounted for by the time dependence of ${\cal
  J}_{\alpha\alpha'}(t)$ with $\alpha\neq\alpha'$.  The dependence
  of the right-hand side of Eq.~(\ref{J}) on the indices $k$ is
  negligible near the Fermi level; therefore we can disregard this
  dependence in the coupling constants ${\cal J}$.

In this paper, we are primarily interested in the irradiation effects
when the applied fields are unable to ionize the dot,
\begin{eqnarray*}
&&eV_{\rm dc},\;eV_{\rm dot},\; eV_{\rm ac} \ll E_d,\;U-E_d\;,\\
&&\hbar\omega,\;\hbar\omega'\ll E_d,\; U-E_d\;.
\end{eqnarray*}
Under these conditions, one can use the approximate solution
(\ref{akalphaadiab}) of the equation (\ref{accondition}), expanding it
in powers of small parameters $eV_{\rm dot}/E_d$, $eV_{\rm
  dot}/(U-E_d)$, {\em etc.}  For simplicity we will consider the cases
when the system is affected by only one kind of ac field: either ac
voltage applied to the gate, or the ac bias.

In the former case, $eV_{\rm dot}\neq 0$, $eV_{\rm ac}=0$ , we obtain
the following expression for the coupling parameters:
\begin{equation}
{\cal J}_{\alpha\alpha'}(t)=
{\cal J}_{\alpha\alpha'}^{(0)}\left[1
+
\gamma\cos\omega t\right]\exp\left[\frac{ie}{\hbar}(V_{{\rm dc},\alpha}-
V_{{\rm dc},\alpha'})t\right]\;,\label{Jreduced}
\end{equation}
where the exchange constants ${\cal J}^{(0)}_{\alpha\alpha'}$ are given by
\begin{equation}
{\cal J}_{\alpha\alpha'}^{(0)}\equiv
\frac{\sqrt{\Gamma_\alpha\Gamma_{\alpha'}}}{\pi\nu\tilde{E}_d},
\quad \tilde{E}_d\equiv\frac{(U-E_d)E_d}{U}\;.\label{J0}
\end{equation}
The exponential factor in Eq.~(\ref{Jreduced}) is due to the dc bias,
which produces the phase difference between the electrons in the left
and right lead. The cosine term accounts for the applied ac field and
stems from the adiabatic variation of the electron energy in the dot,
$E_d+eV_{\rm dot}(t)$, see Eq.~(\ref{akalphaadiab}). The strength of
the applied ac field is characterized by the dimensionless parameter
\begin{equation}
  \gamma\equiv eV_{\rm dot}
\frac{2E_d-U}{(U-E_d)E_d}
\ll 1 \;. 
\label{gamma}
\end{equation}

If the ac field is applied to the leads rather than to the gate,
$eV_{\rm dot}= 0$, $eV_{\rm ac}\neq0$, the expressions for ${\cal
  J}_{\alpha\alpha'}(t)$ read
\begin{eqnarray}
  {\cal J}_{\alpha\alpha}(t)&=&{\cal J}_{\alpha\alpha}^{(0)}\;,
\nonumber\\
{\cal J}_{LR}(t)&=&{\cal J}_{LR}^{(0)}\exp\left[
\frac{ieV_{\rm dc}t}{\hbar}+i\gamma'\sin(\omega't+\phi_0)\right]\;,
  \label{acbiasJfull}
\end{eqnarray}
where ${\cal J}_{\alpha\alpha'}^{(0)}$ is given by Eq.~(\ref{J0}).
The ac bias creates the phase difference between the electrons in the
left and right lead, and therefore enters the exponent in
(\ref{acbiasJfull}) together with the dc bias. The relevant parameter
characterizing the strength of the ac perturbation here is
\begin{equation}
  \label{gammaprim}
  \gamma'\equiv\frac{eV_{\rm ac}}{\hbar\omega'}\;.
\end{equation}
The variation of the electron energy in the dot with respect to the
leads, $E_d\pm eV_{\rm ac}(t)$, see Eq.~(\ref{akalphaadiab}),
generates a term smaller by a factor of $\sim
\hbar\omega'/\tilde{E}_d$, and is neglected in Eq.~(\ref{acbiasJfull}).

In the limit of small amplitude of ac
bias, $\gamma'\ll 1$, the expression
(\ref{acbiasJfull}) for ${\cal
  J}_{\alpha\alpha'}(t)$ may be further simplified by dropping the terms of
high orders in $\gamma'$. Expanding the
right-hand side of (\ref{acbiasJfull}) in
powers of $\gamma'$ up to the first power, we arrive at
\begin{eqnarray}
  {\cal J}_{\alpha\alpha}(t)&=&{\cal J}_{\alpha\alpha}^{(0)}\;,
\nonumber\\
{\cal J}_{LR}(t)&=&{\cal J}_{LR}^{(0)}\exp\left[
\frac{ieV_{\rm dc}t}{\hbar}\right]
\left[1+i\gamma'\sin(\omega't+\phi_0)\right]\;.
  \label{acbiasJ}
\end{eqnarray}

\subsection{Kondo conductance in equilibrium}
\label{sec:3rdorder}

In the framework of the Kondo Hamiltonian (\ref{HK})-(\ref{J}), two
types of tunneling between the left and right leads are possible:
regular elastic cotunneling [the first term in parentheses in
Eq.~(\ref{HKt})], and ``exchange cotunneling'' [the second term]. In
an act of ``exchange cotunneling,'' the simultaneous flip of the spins
of the tunneling electron and of the dot can occur. In the case of
weak coupling ($\nu|{\cal J}_{\alpha\alpha'}^{(0)}|u\ll 1$), one may
apply the perturbation theory to evaluate the conductance through the
dot. It turns out that at $T\to 0$, the higher-order terms of the
perturbation theory series grow, finally making the series diverge,
signaling the Kondo anomaly.  This phenomenon was extensively studied
for the magnetic impurities in metals.\cite{Hewson} In the subsection
\ref{sec:3rdorder} we demonstrate how a similar behavior emerges in
the tunneling through a quantum dot.  The main purpose of this
subsection is to present the formalism which is suitable for the
treatment of a non-equilibrium case at hand. For simplicity, we first
consider the case of no ac field.  Effects of the ac field are included into
consideration in subsequent sections.

Unlike the conventional treatment of the Kondo problem,\cite{Hewson}
we have to consider the Kondo anomaly directly in the conductance,
rather than in the scattering amplitude. This need emerges from the
kinetic nature of the problem at $\gamma,\gamma'\neq0$.  To calculate the
differential dc conductance $G(V_{\rm dc})$, we employ the non-equilibrium
Keldysh technique in the time representation. In this formalism
\begin{equation}
G(V_{\rm dc})=\frac{\partial}{\partial V_{\rm dc}}\langle
{\sf S}(-\infty,0)\hat{I}(0)\, {\sf S}(0,-\infty)
\rangle_0\;,\label{G}
\end{equation}
where
\begin{eqnarray}
\hat{I}(t)&=&\frac{ie}{\hbar}
\sum_{\parbox{0.2in}{\scriptsize
$k$,$\sigma$\\
$k'\!,\sigma'$}}
\left[{\cal J}_{LR}(t)
\left(\textstyle\frac{1}{4}\delta_{\sigma\sigma'}+
\hat{S}_l s^l_{\sigma\sigma'}\right)
c^{\dagger}_{k\sigma L}
c^{\phantom{\dagger}}_{k'\sigma'R}\right.\nonumber\\
&&\left.\phantom{\left(\textstyle\frac{1}{4}\delta_{\sigma\sigma'}+
\hat{S}_l s^l_{\sigma\sigma'}\right)}
-{\rm H.c.}\right]
\label{I}
\end{eqnarray}
is the current operator,
and ${\sf S}(t_2,t_1)$ is the evolution matrix determined by
$\hat{H}_{\cal J}$.

In the lowest non-vanishing (second) order of the perturbation theory
in the coupling constant ${\cal J}^{(0)}_{\alpha\alpha'}$, the
conductance of the dot is given by the expression
\begin{equation}
  \label{G2}
  G^{(2)}=\pi^2 \frac{e^2}{\pi\hbar}\nu^2\left[{\cal J}_{LR}^{(0)}\right]^2\;.
\end{equation}

The logarithmic divergences appear starting from the terms of the
third order in ${\cal J}_{\alpha\alpha'}^{(0)}$. A representative term
has the following structure:
\begin{eqnarray}
&& \frac{e^2}{\pi\hbar}
\frac{\left[{\cal J}_{LR}^{\rm (0)}\right]^2
{\cal J}_{RR}^{\rm (0)}}{\hbar^3}
\int_{-\infty}^0 dt_1 \int^0_{t_1} dt_2
\langle \hat{S}_j(0)  \hat{S}_k(t_1)  \hat{S}_l(t_2) \rangle
\varepsilon^{jkl}\nonumber\\
&&\qquad\times
\left[t_1 \cos (eVt_1/\hbar) +
t_2\cos (eVt_2/\hbar)\right]\nonumber\\
&&\qquad\times\sum_{k_1,k_2,k_3}G_{k_1}(-t_2)
G_{k_2}(t_2-t_1)\bar{G}_{k_3}(t_1)
\;,\label{typterm}
\end{eqnarray}
$\varepsilon^{jkl}$ is the antisymmetric unit tensor and $G_k(t)$ and
$\bar{G}_k(t)$ are the time-ordered and anti-time-ordered
Green functions of free electrons in the leads, given by
\begin{equation}
G_k(t)=\left\{
\begin{array}{ll}
-i[1-f(\xi_k)]\;,\;&\mbox{if}\quad t>0\;,\\
i\,f(\xi_k)]\;,&\mbox{if}\quad t<0\;,
\end{array}
\right.
\end{equation}
with $f(\xi)$ being the Fermi distribution function. This and
other terms of the same structure yield the Kondo divergence in the
conductance.

If there is no external ac field, the averages $\langle \hat{S}_j(t_1)
\hat{S}_k(t_2) \hat{S}_l(t_3) \rangle$ are independent of time and
equal $(i/4)\varepsilon_{jkl}$.  After adding up all the cubic in
${\cal J}^{\rm (0)}_{\alpha\alpha'}$ terms in the expression for the
conductance $G$ [one of them is given by Eq.~(\ref{typterm})], summing
over the electron states $k_i$, and performing the integration over
$t_2$ [see Eq.~(\ref{typterm})], we arrive at
\begin{eqnarray}
&&G^{(3)}(T,V_{\rm dc})=\frac{3\pi^2}{2} \frac{e^2}{\pi\hbar}
\nu^3\left[{\cal J}_{LR}^{\rm (0)}\right]^2
\left[{\cal J}_{RR}^{\rm (0)}+{\cal J}_{LL}^{\rm (0)}\right]
\nonumber\\
&&\quad
\times\int^0_{-\infty}\!\! dt\,
\frac{(-t)\cos (eV_{\rm dc}t/\hbar)}
{\sinh^2(\pi Tt/\hbar)+(T/D_0)^2}\left(\frac{\pi T}{\hbar}\right)^2
\label{Gt}
\end{eqnarray}
Here 
\begin{equation}
  \label{D0}
 D_0\equiv \sqrt{E_d(U-E_d)} 
\end{equation}
is the effective bandwidth.\cite{Haldane} For the sake of simplicity,
further we will mostly consider the zero-bias conductance $G_{\rm
  peak}$. In this case, Eq.~(\ref{Gt}) yields
\begin{equation}
  \label{G3}
  G^{(3)}_{\rm peak}(T)=\frac{3\pi^2}{2} \frac{e^2}{\pi\hbar}
\nu^3\left[{\cal J}_{LR}^{\rm (0)}\right]^2
\left[{\cal J}_{RR}^{\rm (0)}+{\cal J}_{LL}^{\rm (0)}
\right]\ln\frac{D_0}{T}\;.
\end{equation}
The results for the finite-bias conductance $G^{(3)}(V_{\rm dc})$ with
$eV_{\rm dc}>T$ can be obtained from Eq.~(\ref{G3}) by replacing $T$
with $eV_{\rm dc}$.

Thus the second [Eq.~(\ref{G2})] and third [Eq.~(\ref{G3})] orders of
the perturbation theory in the coupling constant ${\cal
  J}_{\alpha\alpha'}^{(0)}$ yield the following expression for the dot
conductance
\begin{eqnarray}
  G_{\rm peak}&=&\frac{3\pi^2}{4}
\frac{e^2}{\pi\hbar}\nu^2\left[{\cal J}_{LR}^{(0)}\right]^2
\left[1+2\nu\left({\cal J}_{RR}^{\rm (0)}+{\cal J}_{LL}^{\rm
(0)}\right)\ln\frac{D_0}{T}\right]\nonumber\\
&+&\frac{\pi^2}{4} \frac{e^2}{\pi\hbar}\nu^2\left[{\cal J}_{LR}^{(0)}\right]^2\;.
  \label{G23}
\end{eqnarray}
Here we have split the quadratic in ${\cal J}^{(0)}_{\alpha\alpha'}$
contribution (\ref{G2}) in two: the one due to the ``exchange
cotunneling'', which entered the first term in Eq.~(\ref{G23}), and
the one due to regular cotunneling, which became the last term in
Eq.~(\ref{G23}). The cubic in ${\cal J}^{(0)}_{\alpha\alpha'}$ term in
Eq.~(\ref{G23}) grows as the temperature is being lowered,
demonstrating the Kondo anomaly. The regular cotunneling does not
produce terms growing at low temperatures and bias, and does
not contribute to the Kondo effect. Equation (\ref{G23}) is valid
while $T\gg T_K\sim D_0\exp[1/\nu ({\cal
  J}^{(0)}_{LL}+{\cal J}^{(0)}_{RR})]$.

If this condition  is not satisfied, then the
expansion up to the cubic in ${\cal J}_{\alpha\alpha'}^{(0)}$ terms is
insufficient.  At $T\gtrsim T_K$, the conductance can be derived in
the leading logarithmic approximation.  The latter consists in the
summation of the most diverging terms in each order in ${\cal
  J}^{(0)}_{\alpha\alpha'}$, {\em i.e.} terms, proportional to $[{\cal
  J}^{(0)}_{LR}]^2[{\cal J}^{(0)}_{\alpha\alpha'}
\ln(D_0/T)]^n$, in the series for $G$. To perform this summation, we
modify the ``poor man's scaling'' technique.\cite{Anderson} In the
framework of this technique, the electron bandwidth $D$ is gradually
reduced, and the exchange constants in the Kondo Hamiltonian
(\ref{HK}) are renormalized to compensate for this band reduction,
{\em i.e.}  ${\cal J}_{\alpha\alpha'}^{(0)}$ is replaced with some
${\cal J}_{\alpha\alpha'}(D)$ . The proper dependence of ${\cal
  J}_{\alpha\alpha'}$ on $D$ should be derived from the condition of
invariance of physical quantities with respect to the RG
transformation.  Finally, the renormalized Hamiltonian with the
reduced band width will allow for the calculation of the conductance
in the second order of the perturbation theory in the renormalized
exchange constants ${\cal J}_{\alpha\alpha'}$; the resulting
expression will be equal to the sum of the dominant terms of all
orders of the perturbation theory in the initial, bare exchange
constants ${\cal J}^{(0)}_{\alpha\alpha'}(D=D_0)$.

For the non-equilibrium system we consider, the Renormalization Group
(RG) equations for the exchange constants should be derived from the
condition of the invariance of the linear conductance (or current)
under the RG transformation, rather than the invariance of the
scattering amplitudes. In the main logarithmic approximation which we
are going to employ, the (invariant) conductance must be evaluated in
the two lowest non-vanishing orders of the perturbation theory, namely
in the second and third ones, see Eqs.~(\ref{G2}), (\ref{Gt}).  The
Kondo divergence (and, therefore, the renormalization of ${\cal
  J}_{\alpha\alpha'}$) occur due to exchange scattering [the second
term in braces in Eq.~(\ref{HKt})] only.  Therefore we single out this
contribution in the term of the second order in ${\cal
  J}_{\alpha\alpha'}$,
\begin{equation}
G_{\rm exch}^{(2)}(D)=\frac{3\pi^2}{4} \frac{e^2}{\pi\hbar} \nu^2
\left[{\cal J}^{(0)}_{LR}(D)\right]^2\;.
\label{I2}
\end{equation}
In the third order in the exchange constants, the conductance is
given by Eq.~(\ref{Gt}).  The resulting condition of invariance of $G$
under the transformation, which corresponds to the ``poor man`s
scaling'', has the following form:
\begin{eqnarray}
\frac{\partial}{\partial D}&&\left\{G_{\rm exch}^{(2)}(D)\right.
\nonumber\\
&&\quad\left.+
\frac{3\pi^2}{2} \frac{e^2}{\pi\hbar}\nu^3 \left[{\cal J}_{LR}\right]^2
\left[{\cal J}_{RR}+{\cal J}_{LL}\right] 
\ln\frac{D}{T}
\right\}=0\label{RGeq}\;.
\end{eqnarray}
Within the accuracy of this equation, when differentiating the second
term, we should neglect any implicit dependence on $D$ through the
parameters ${\cal J}_{\alpha\alpha'}(D)$. 

Equation (\ref{RGeq}), together with Eq.~(\ref{I2}), yields the
equation for the evolution of ${\cal J}_{LR}$:
\begin{equation}
  \label{RGeq1}
  \frac{d{\cal J}_{LR}}{dD}=\nu
\frac{{\cal J}_{LR}({\cal J}_{RR}+{\cal J}_{LL})}{D}\;,
\end{equation}
The corresponding equations for ${\cal J}_{RR}$ and ${\cal J}_{LL}$
can be derived from the condition of invariance of other physical
quantities under the RG transformation.  For this purpose, we pick the
spin current from the left and right lead,
\begin{eqnarray}
  \label{spincur}
  I_{\alpha}^{(s)}&=&
\langle {\sf S}(-\infty,0) \hat{I}_{\alpha}^{(s)}(0) {\sf S}(0,-\infty)
\rangle_0\;,\\
\hat{I}_{\alpha}^{(s)}(t)&\equiv&i\left[\hat{H}_{\cal J}, 
\sum_k\left(c^\dagger_{k\uparrow\alpha} c^{\phantom{\dagger}}_{k\uparrow\alpha}-
c^\dagger_{k\downarrow\alpha}
c^{\phantom{\dagger}}_{k\downarrow\alpha}
\right) \right]\;,
\end{eqnarray}
which is induced by applying infinitely small magnetic field to the leads. 
The resulting two equations will be independent, in contrast to the
corresponding equations for the charge, because the spin of the dot
can vary while the charge cannot.

Evaluating $I_{\alpha}^{(s)}$ in the second and third orders of the
perturbation theory in ${\cal J}^{(0)}_{\alpha\alpha'}$, similarly to
Eq.~(\ref{G23}), and differentiating it by $D$, we arrive at
\begin{eqnarray}
  \label{RGeq2}
  \frac{d{\cal J}_{RR}}{dD}=\nu
\frac{{\cal J}^2_{RR}+{\cal J}_{LR}^2}{D}\;,\\
  \label{RGeq3}
  \frac{d{\cal J}_{LL}}{dD}=\nu
\frac{{\cal J}^2_{LL}+{\cal J}_{LR}^2}{D}\;,
\end{eqnarray}
Equations (\ref{RGeq1}), (\ref{RGeq2}), and (\ref{RGeq3}) make a
complete system, which, with the initial conditions
\begin{equation}
  \label{RGinit}
{\cal J}_{\alpha\alpha'}(D_0)=
{\cal J}_{\alpha\alpha'}^{(0)}\equiv
\frac{\sqrt{\Gamma_\alpha\Gamma_{\alpha'}}}{\pi\nu\tilde{E}_d}\;,  
\end{equation}
see Eq.~(\ref{J0}), yields
\begin{equation}
  {\cal
    J}_{LR}(D)=\frac{2\sqrt{\Gamma_L\Gamma_R}}{\Gamma_L+\Gamma_R}
    \frac{1}{2\nu\ln(D/T_K)}\;. 
  \label{RGsolut}
\end{equation}
The Kondo temperature $T_K$ here is given by
\begin{equation}
T_K=\mu \sqrt{\frac{(\Gamma_L+\Gamma_R) U}{\pi}}
\exp\left[-\frac{\pi \tilde{E}_d}{(\Gamma_L+\Gamma_R)}\right]\;,
\label{TK}
\end{equation}
with $\mu\sim 1$. To obtain the pre-exponential factor
$\sqrt{(\Gamma_L+\Gamma_R) U/\pi}$ in the equation for $T_K$ one has
in fact to include the next in ${\cal J}_{\alpha\alpha'}(D)$ order in the RG
equations, see Ref.~\onlinecite{Haldane}.

The renormalization should proceed until the band width is reduced to
$T$. After that, the current and conductance can be calculated in the
Born approximation (\ref{I2}) in the renormalized exchange constant
${\cal J}_{LR}$ given by Eq.~(\ref{RGsolut}) with $D=T$.  The
resulting expression for the conductance in the domain $T\gtrsim T_K$
is
\begin{equation}
G_{\rm peak}=
\frac{3\pi^2}{16}
\frac{1}{[\ln (T/T_K)]^2}G_U\;,
\label{GRG}
\end{equation}
where 
\begin{equation}
  \label{GU}
  G_U\equiv \frac{e^2}{\pi\hbar}
  \frac{4\Gamma_L\Gamma_R}{(\Gamma_L+\Gamma_R)^2}\;
\end{equation}
is the conductance of the dot in the unitary limit of
tunneling.

At $T\gg T_K$, one can expand Eq.~(\ref{GRG}) into the series of
powers of ${\cal J}_{\alpha\alpha'}^{(0)}\ln(D_0/T)$. The first term
of the series is the conductance calculated in the Born approximation
[see Eq.~(\ref{I2})], the second term yields the lowest order Kondo
correction given by Eq.~(\ref{G3}).

The RG technique can be also used to derive the dependence of Kondo
conductance on the applied dc bias in the domain $eV_{\rm dc}\gtrsim
T_K$, $eV_{\rm dc}> T$.  Starting from Eqs.~(\ref{Gt}) and (\ref{I2})
and proceeding along the lines of Eqs.~(\ref{RGeq})--(\ref{RGsolut}),
we arrive at
\begin{equation}
G(V_{\rm dc})=
\frac{3\pi^2}{16}
\frac{1}{[\ln (eV_{\rm dc}/T_K)]^2}G_U\;.
\label{GRGV}
\end{equation}

Thus the Renormalization Group technique (\ref{RGeq})--(\ref{GRG})
allows one to perform summation of infinite series of the perturbation
theory in the exchange constants ${\cal J}_{\alpha\alpha'}^{(0)}$. The
results obtained in this way are valid in a wider domain of parameters
as compared to the results of the finite-order perturbation theory.
The RG technique reveals the meaning of the energy scale $T_K$. The
resulting expressions (\ref{GRG}),~(\ref{GRGV}) for physical
quantities contain the single relevant characteristic of the system,
$T_K$, rather than numerous parameters of the Anderson Hamiltonian
[Eq.~(\ref{H})].  For example, in Eq.~(\ref{GRG}) the dependence of
the differential conductance on the applied bias is expressed in terms
of the dimensionless variable $T/T_K$. The dependence of
$G/G_U$ on this variable is given by some universal function
of any value of $T/T_K$;\cite{Hewson} its high-temperature
asymptote (\ref{GRG}) is established with the help of RG technique.
Similarly, the frequency and magnitude of the applied ac field may
enter into some new universal formulas for $G/G_U$ in the form of
dimensionless variables, being normalized by $T_K$. The generalization
of the RG technique which we presented in this section, will allow us
to check the validity of this conjecture and to establish the
asymptotes of these new universal dependences.

\section{Spin decoherence by ac gate voltage}
\label{sec:weak}

Now we include into consideration the effects of an ac field.  As we
have shown in our earlier paper,\cite{KaminskiEtal99} the ac field can
bring decoherence into the dynamics of the dot's spin, thus affecting
the Kondo conductance.  We start our study of the irradiation-modified
Kondo anomaly from the consideration of the decoherence.

\subsection{Mechanisms of spin decoherence}\label{sec:decoh}

In terms of the Anderson Hamiltonian (\ref{H}) the loss of
coherence by the dot spin occurs when an electron leaves the dot and
another electron, with the opposite spin, enters it. If the frequency
of the applied ac field is large enough, $\hbar\omega>E_d,U-E_d$, this
process can consist of two real processes: the dot gets ionized by the
ac field, and then an electron from a lead enters the dot to fill the
vacancy. Alternatively, an extra electron can be put in the dot, and
then the electron which was initially present in the dot leaves it.

\begin{figure}
\epsfxsize=8.5cm
\centerline{\epsfbox{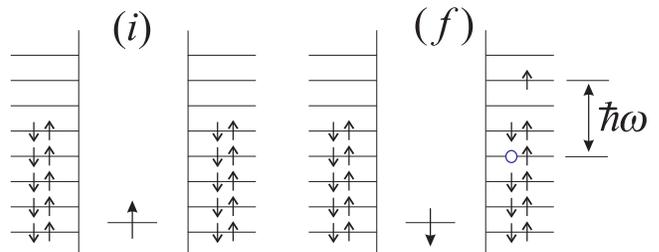}}
\caption{\label{fig:transAC}
{\bf Spin-flip cotunneling}: 
absorbing a photon, an electron hops from a state below the Fermi
level to a state above the Fermi level; the spins of the electron and
of the dot flip due to exchange interaction between them.
}
\end{figure}

In the present paper we deal with a more subtle case, when the applied
ac field is unable to ionize the dot.  In this case the dot can still
change its spin, even at zero bias, by means of the ``spin-flip
cotunneling,'' which is shown schematically on Fig.~\ref{fig:transAC}.
In the course of this process, an electron, which interacts with the
dot spin [see Eq.~(\ref{HK})], absorbs a photon and hops to a state
above the Fermi level, while the spin of the dot flips.  In terms in
the Anderson Hamiltonian (\ref{H}), this process cannot be described
as two separate real processes. Instead, the change of the dot spin
occurs as a single process when a state with two or no electrons in
the dot appears only as a virtual intermediate state. 

\narrowtext

The rate of the spin-flip cotunneling can be calculated with the help
of the Kondo Hamiltonian given by Eqs.~(\ref{HK}), (\ref{J}), and
(\ref{akalphaadiab}). In the case of weak modulation, $\gamma\ll 1$,
see Eq.~(\ref{acbiasJfull}), it is sufficient to account for the
single-photon processes only, and use the reduced form of the
Hamiltonian, given by Eqs.~(\ref{HK}), (\ref{Jreduced})-(\ref{gamma}).
An infinitely small dc bias, needed for measurements of the linear
conductance, does not affect the rate of spin-flip cotunneling.
Therefore in this subsection we set $V_{\rm dc}=0$ for the sake of
simplicity.

Applying the Fermi Golden Rule we obtain
\begin{equation}
\frac{\hbar}{\tau}=\frac{1}{8\pi}\hbar\omega
\left[\frac{\Gamma_L+\Gamma_R}{\tilde{E}_d}\right]^2
\gamma^2\;,
\label{tau}
\end{equation}
where $\gamma$ is given by Eq.~(\ref{gamma}).

The amplitude of inelastic transitions yielding Eq.~(\ref{tau}) was
evaluated in the lowest-order perturbation theory. It corresponds to
the first order in the amplitude of the ac perturbation, and the
zeroth order in the time-independent (at $V_{\rm dc}=0$) part of the
exchange interaction (\ref{HKt}). Accounting for the terms of
higher-order in this time-independent part renormalizes the amplitude
of the inelastic transition (which is still linear in the amplitude of
the ac field). Similarly to the calculation of the conductance, we
intend to collect the leading logarithmic terms in the renormalization
of the inelastic transition amplitude. This can be accomplished by the
renormalization group (RG) transformation, described in
subsection \ref{sec:3rdorder}. The transformation reduces the electron
band width $D$ and renormalizes the matrix elements ${\cal
  J}_{\alpha\alpha'}$ of the Kondo Hamiltonian (\ref{HK}) to account
for this band reduction. Finally one can calculate the decoherence
rate in the second order perturbation theory in renormalized ${\cal
  J}_{\alpha\alpha'}$; the result given by such a treatment equals the
sum of infinite perturbation theory series in the initial Hamiltonian.

The RG transformation starts from the bandwidth $D=D_0$ given by
Eq.~(\ref{D0}), and the initial matrix elements 
\begin{equation}
  \label{tauRGJ}
  \left.{\cal J}_{\alpha\alpha'}(D)\right|_{D=D_0}=
{\cal J}_{\alpha\alpha'}^{(0)}\left[1
+\gamma\cos\omega t\right]\;,
\end{equation}
cf. Eq.~(\ref{Jreduced}).  While the width $D$ of the band exceeds
$\hbar\omega$, the time dependence of the Hamiltonian matrix
elements~(\ref{tauRGJ}) can be treated adiabatically, {\em i.e.}  time
$t$ in the right-hand side of Eq.~(\ref{tauRGJ}) can be considered as
just a parameter. The RG equations, derived from the condition of
invariance of physical quantities under the transformation, have the
same form (\ref{RGeq1}), (\ref{RGeq2}), and (\ref{RGeq3}). The
transformation must be stopped when the bandwidth is reduced to the
values of the order of the frequency $\hbar\omega$ of the applied ac
field. Expanding the solution of the RG equations (\ref{RGeq1}),
(\ref{RGeq2}), and (\ref{RGeq3}) with the initial condition
(\ref{tauRGJ}) in powers of $\gamma$ up to the first power, we obtain
\begin{eqnarray}
\left.{\cal J}_{\alpha\alpha'}(D)\right|_{D\sim\hbar\omega}&=&
\frac{2\sqrt{\Gamma_\alpha\Gamma_{\alpha'}}}{\Gamma_\alpha+\Gamma_{\alpha'}}
    \frac{1}{2\nu\ln(\hbar\omega/T_K)}\nonumber\\
&\times&\left[1+
\gamma\frac{\pi\tilde{E}_d}{\Gamma_L+\Gamma_R}  
  \frac{1}{\ln(\hbar\omega/T_K)}\cos\omega t\right]
\;.
  \label{RGsolutprim}
\end{eqnarray}
The Fermi Golden rule applied to the Hamiltonian (\ref{HK}) with
${\cal J}_{\alpha\alpha'}$ given by Eq.~(\ref{RGsolutprim}) yields the
following expression for the decoherence rate:
\begin{equation}
  \label{tauu}
  \frac{\hbar}{\tau T_K}=\frac{3\pi}{32}
  \frac{\hbar\omega}{T_K}
  \frac{1}{\left[\ln(\hbar\omega/T_K)\right]^4}
  \left[\frac{\delta T_K}{T_K}\right]^2\;.
\end{equation}
Here we introduced the relative amplitude 
\begin{equation}
  \label{deltaTK}
\frac{\delta T_K}{T_K}\equiv
  \gamma \frac{\pi
\tilde{E}_d}{\Gamma_L+\Gamma_R}\;
\end{equation}
of adiabatic variations of the
``time-dependent Kondo temperature''. The latter is defined by
\begin{equation}
  \label{TKt}
  T_K(t)\equiv\mu \sqrt{\frac{(\Gamma_L+\Gamma_R) U}{\pi}}
\exp\left[-\frac{\pi \tilde{E}_d(t)}{(\Gamma_L+\Gamma_R)}\right]\;,
\end{equation}
with 
\begin{equation}
  \label{tildeEdt}
\tilde{E}_d(t)=
\tilde{E}_d(1+\gamma\cos\omega t)\;,
\end{equation}
cf. Eqs.~(\ref{TK}) and (\ref{J0}). 

One can see that the amplitude of the ac field enters
Eq.~(\ref{tauu}) through the dimensionless parameter $\delta T_K /T_K$. The
value of this parameter, in principle, can be directly measured. 
Representation of $\hbar/\tau T_K$ in terms of $\delta T_K /T_K$ will
allow us to build a universal description of the ac field effect on
the Kondo conductance.

As we mentioned before, the spin-flip cotunneling is essentially
different from the dot ionization with its subsequent filling. During
a process of spin-flip cotunneling, the ionized dot exists only as a
virtual state.  Therefore, the spin-flip cotunneling persists at
frequencies $\min\{E_d,U-E_d\}>\hbar\omega$, leading to the 
decoherence of the dot spin state without ionization.

\subsection{Effects of spin decoherence on Kondo conductance}
\label{sec:AffectedWeakKondo}

As we have just shown, the external ac field is able to flip the dot's
spin. Therefore, in the presence of the ac field, the averages of the
type $\langle S_j(t_1) S_k(t_2) S_l(t_3)\rangle$ no longer equal
$\langle S_j(0) S_k(0) S_l(0)\rangle=\mathop{{\rm Tr}} S_j S_k S_l
\equiv (i/4) \varepsilon_{jkl}$.  In the limiting case
$|t_m-t_n|\gg\tau$ ($m\neq n$), the orientations of the dot spin at
$t=t_1$, $t_2$, $t_3$ are independent of each other, because of the
spin-flip cotunneling, and one has
\begin{displaymath}
  \langle S_j(t_1) S_k(t_2) S_l(t_3)\rangle=
  \langle S_j(t_1) \rangle \langle S_k(t_2) \rangle \langle S_l(t_3)\rangle=0\;.
\end{displaymath}
At finite time intervals $|t_m-t_n|$, the spin correlator decays
exponentially, with the spin-flip cotunneling rate  being the
characteristic decay rate:
\begin{eqnarray}
\langle \hat{S}_j(t_1) \hat{S}_k(t_2) \hat{S}_l(t_3) \rangle=(i/4)
\varepsilon_{jkl}
\exp(-t_{max}/\tau)\;,
\label{SSS}\\
t_{max}\equiv\max\{|t_1-t_2|, |t_2-t_3|,
|t_1-t_3|\}\;.
\nonumber
\end{eqnarray}
Equation (\ref{SSS}) can be derived using the formalism of
equations of motion.  In the framework of this formalism,
Eq.~(\ref{SSS}) appears as the solution of the equation
\begin{eqnarray}
&&  \frac{\partial}{\partial t_1}\langle \hat{S}_j(t_1) \hat{S}_k(t_2)
  \hat{S}_l(t_3) \rangle\nonumber\\
&&\quad=\langle {\sf S}(t_3,t_1) \left(i[\hat{H}_{\cal J},\hat{S}_j]\right){\sf
  S}(t_1,t_2) \hat{S}_k {\sf S}(t_2,t_3)\hat{S}_l \rangle_0 
\label{em1}
\end{eqnarray}
where ${\sf S}(t,t')$ is the evolution matrix determined by
$\hat{H}_{\cal J}$. Expanding ${\sf S}(t,t')$ in powers of
$\gamma{\cal J}_{\alpha\alpha'}^{(0)}$ up to the first power, we arrive at
\begin{eqnarray}
&&  \frac{\partial}{\partial t_1}\langle \hat{S}_j(t_1) \hat{S}_k(t_2)
  \hat{S}_l(t_3) \rangle\nonumber\\
&&\quad =\frac{1}{\tau}\left[\theta(t_3-t_1)\theta(t_2-t_1)-
\theta(t_1-t_3)\theta(t_1-t_2)\right]\nonumber\\
&&\qquad\times\langle \hat{S}_j(t_1) \hat{S}_k(t_2)
  \hat{S}_l(t_3) \rangle\;,\label{em}
\end{eqnarray}
where $\tau$ is given by Eq.~(\ref{tau}). Equation (\ref{em}) with
$\tau$ given by Eq.~(\ref{tauu}) can be obtained by expanding the
evolution matrix ${\sf S}(t,t')$ up to the second power in
$\gamma{\cal J}_{\alpha\alpha'}^{(0)}$, and using the RG technique
described in Sec.~\ref{sec:3rdorder}. 

The leading effect of the irradiation is in cutting off the
logarithmic divergences in the time integrals of type
~(\ref{typterm}). One can easily see that with the time-decaying spin
correlation function (\ref{SSS}), $G^{(3)}_{\rm peak}$ is finite even at $T\to
0$:
\begin{equation}
  \label{newG3}
G^{(3)}=\frac{3\pi^2}{2} \frac{e^2}{\pi\hbar}
\nu^3\left[{\cal J}_{LR}^{\rm (0)}\right]^2
\left[{\cal J}_{RR}^{\rm (0)}+{\cal J}_{LL}^{\rm
(0)}\right]\ln\frac{D_0\tau}{\hbar}\;.
\end{equation}
As we have shown in Sec.~\ref{sec:decoh}, the spin decoherence by
external irradiation does not require ionization of the impurity
level, and therefore exists at frequencies below $E_d$, $U-E_d$.

The effect of  the irradiation on the Kondo conductance  is
not analytic in the intensity of the ac field. It cannot be obtained
by a finite-order perturbation theory in $\gamma$ in the formula (\ref{G}).
To obtain Eq.~(\ref{newG3}) directly from Eq.~(\ref{G}) using the
perturbation theory series in ${\cal J}_{\alpha\alpha'}^{(0)}$, one
would need to add up all the terms proportional to $[{\cal
  J}_{\alpha\alpha'}^{(0)}]^3[\gamma{\cal J}_{\alpha\alpha'}^{(0)}]^{2n}$.

The finite-order perturbation theory [Eqs.~(\ref{G2}) and
(\ref{newG3})] can be used to evaluate the Kondo conductance only if
the decoherence rate $\hbar/\tau$ is much larger than the Kondo
temperature $T_K$. At lower decoherence rates we have to take into
account terms of all orders in ${\cal J}_{\alpha\alpha'}^{(0)}$. It
can be done by means of the Renormalization Group technique described
in Sec.~\ref{sec:3rdorder}. One RG equation is to be derived from the
condition of invariance of the conductance, given by the second and
third orders of the perturbation theory in ${\cal
  J}_{\alpha\alpha'}^{(0)}$ [Eqs.~(\ref{I2}) and (\ref{newG3})],
similarly to Eq.~(\ref{RGeq1}). The other two RG equations can be
obtained using the requirement of invariance of the spin current
(\ref{spincur}) under the RG transformation. The resulting set of
equations coincides with the one given by
Eqs.~(\ref{RGeq1}),~(\ref{RGeq2}), and~(\ref{RGeq3}).  When the
decoherence rate exceeds the temperature $T$, the RG transformation
must be stopped when the bandwidth $D$ reaches $\hbar/\tau$ rather
than $T$.  Then the linear conductance can be evaluated in the second
order perturbation theory in the renormalized exchange constants
${\cal J}_{LR}$, given by Eq.~(\ref{RGsolut}) with $D=\hbar/\tau$:
\begin{equation}
G_{\rm peak}=
\frac{3\pi^2}{16}
\frac{1}{\left[\ln(\hbar/\tau T_K)\right]^2}
G_U\;.
\label{GRGtau}
\end{equation}
Here $\hbar/\tau T_K$ is determined by Eq.~(\ref{tauu}).  Equation
(\ref{GRGtau}) is the central formula of this section.  It defines the
conductance of the quantum dot as a function of two dimensionless
parameters: $\hbar\omega/T_K$ and $\delta T_K/T_K$
[Eq.~(\ref{deltaTK})].  The region of validity of Eq.~(\ref{GRGtau})
is determined by the condition
\begin{equation}
  \label{weakcond}
  \frac{\hbar}{\tau T_K}\gtrsim 1\;,
\end{equation}
and corresponds to the regime of strong suppression of the Kondo
effect by the external ac field.  At fixed strength of the ac field
the spin-flip rate (\ref{tau}) decreases with the decrease of ac field
frequency $\omega$. Correspondingly, the peak conductance
(\ref{GRGtau}) grows.  The crossover from weak to strong [$G\sim G_U$]
Kondo effect occurs when $\hbar/\tau\sim T_K$. Equations (\ref{tau})
and (\ref{tauu}) show that this value of $\hbar/\tau$ is reached while
$\hbar\omega/T_K\gg 1$. 

\section{Weak spin decoherence}
\label{sec:crossover}

In this section we consider the regime of ``intermediate suppression'' of
the Kondo effect by the ac radiation. By ``intermediate'' we mean that
the decoherence is relatively weak
\begin{mathletters}
  \label{intercond}
  \begin{equation}
    \label{ic1}
\hbar/\tau< T_K \;,    
  \end{equation}
and the formula (\ref{GRGtau}) is no longer valid,
but the frequency is still sufficiently high
  \begin{equation}
  \label{ic2}
  T_K<    \hbar\omega \;,
  \end{equation}
\end{mathletters}
so that the RG result (\ref{tauu}) for the decoherence rate holds.  In
this regime, the formation of the many-body state is not suppressed,
because of Eq.~(\ref{ic1}). However, Eq.~(\ref{ic2}) allows for sudden
spin flips. The complicated nature of the many-body state hampers the
quantitative consideration of this regime, and we limit ourselves to
qualitative analysis.

When the many-body Kondo resonance is fully formed, the conductance of
the dot equals $G_U$ [Eq.~(\ref{GU})] and corresponds to the unitary
limit of tunneling through the dot.  An act of spin flip destroys the
many-body state, and the conductance drops substantially below the
value given by Eq.~(\ref{GU}). The time necessary for the many-body
state to be restored equals approximately
$\hbar/T_K$.\cite{Nordlander99} Therefore, the fraction of time which
the system spends in the highly-conducting ($G\approx G_U$) state
equals approximately $1-a\hbar/\tau T_K$, where $a\sim 1$.  The
resulting time-averaged conductance of the dot can be estimated as
\begin{equation}
  \label{Gcross}
  G_{\rm peak}=\left[1-a\frac{\hbar}{\tau T_K}\right] G_U\;.
\end{equation}
The rate $\hbar/\tau$ of the spin-flip processes here is given by
Eq.~(\ref{tauu}).  Under conditions (\ref{intercond}), parameter $a$
does not depend on the characteristics of the ac field. The value of
$a$ should be found from the quantum-mechanical problem of evolution,
which starts with a state ``prepared'' by the flip of the impurity
spin, and results eventually in the re-formation of a Kondo polaron.
Our qualitative treatment of the regime (\ref{intercond}) does not
allow us to find the exact value of the universal coefficient $a$,
which however can be found from a numeric calculation.

At the upper limit of applicability, $\hbar/\tau\sim T_K$, the peak
conductance given by Eq.~(\ref{Gcross}) matches the result (\ref{GRGtau}).

\section{Low-frequency ac field: adiabatic approximation}
\label{sec:strong}

In Sections \ref{sec:3rdorder} and \ref{sec:weak} we considered the
case when the isolated spin is only weakly screened by the
many-electron state formed around it. The complete screening was
suppressed either by relatively high temperature, $T>T_K$, or by large
bias $eV_{\rm dc}>T_K$, or by the decoherence. In the case of
Sec.~\ref{sec:crossover}, the spin-screening cloud is able to form;
however, the  spin flips, produced by the irradiation, occasionally
destroy this many-body state, thus reducing the conductance. 

In this section, we consider the case of low frequencies of the ac
field $\hbar\omega\ll T_K$, when the energy of a photon is insufficient
to flip the dot's spin in the fully formed many-body Kondo state. For
the irradiation to be the leading cause of the deviation of the
conductance from the unitary limit, we suppose the temperature and
bias to be also low, $T, eV_{\rm dc} \ll T_K$. The RG technique we
used before is not applicable in this regime. Therefore we need
another approach to evaluate the conductance of the quantum dot system
and the effects of the external irradiation on it.

The required approach is provided by the scaling theory of
Nozi\`{e}res and Blandin.\cite{NozieresBlandin80} This theory states
that the renormalization-group transformation, whose initial stage was
described in Sec.~\ref{sec:3rdorder}, can be continued, and finally
leads to a fixed point. At the fixed point, the system exhibits
Fermi-liquid behavior, and its Hamiltonian has a relatively simple
form.\cite{AffleckLudwig93} This fixed-point Hamiltonian can be used
to study the properties of the Kondo system at low temperatures, $T\ll
T_K$.\cite{Nozieres,AffleckLudwig93} Mapping the quantum dot system in
the Kondo regime onto the regular one-channel Kondo problem, we can
employ the fixed-point Hamiltonian to evaluate the dc current through
the dot induced by the applied bias.

The external ac field disturbs the many-particle state formed near the
isolated spin, leading to the deviations of the system behavior from
that dictated by the (static) fixed-point Hamiltonian. In this section
we study the case when the frequency of the field is low
($\hbar\omega\ll T_K$), so that the many-body state is not destroyed
but rather adiabatically varied by the ac field, as the level in the
dot goes up and down [see Eq.~(\ref{H})]. Then the current through the
dot can be evaluated with the help of the fixed-point Hamiltonian with
time-dependent parameters.

Now we map the problem of transport through the dot onto the regular
scattering problem.  For this purpose, it is convenient to use the
basis of $s$ and $p$ scattering states rather than that of the
left-lead and right-lead states.  These two bases are connected by
\begin{eqnarray}
&&a_{k\sigma}^{(s)}=\xi c_{k\sigma L} + \eta c_{k\sigma R}\;,\quad
a_{k\sigma}^{(p)}=-\eta c_{k\sigma L} + \xi c_{k\sigma R}\;,\label{LRtosp}\\
&&\mbox{where}\quad\xi\equiv \frac{v_L}{\sqrt{v_L^2+v_R^2}}\;,
\quad\eta\equiv \frac{v_R}{\sqrt{v_L^2+v_R^2}}\;.
\nonumber
\end{eqnarray}
The $p$-states are decoupled from the dot, so the dot-lead coupling term
in the Anderson Hamiltonian (\ref{H}) has the form
\begin{displaymath}
\sqrt{v_L^2+v_R^2}\,\sum_{k,\sigma}\left(
a_{k\sigma}^{(s)\dagger}d^{\phantom{\dagger}}_{\sigma}+
{\rm H.c.}
\right)\;.
\end{displaymath}
The initial basis $c_{k\sigma\alpha}$ is composed of the states
residing entirely in the left or right lead, which is convenient for
the problem of two leads connected by a weak link, when the inter-lead
tunneling is to be considered as a perturbation. In terms of incident
and reflected/transmitted waves, these states correspond to the waves
incident from one of the leads to the dot and completely reflected back
to the same lead. Therefore the $s$-waves of Eq.~(\ref{LRtosp}), which
enter the new basis, have the scattering phase equal to $\pi/2$.

Making the Schrieffer-Wolff transformation, we arrive to the regular
Kondo problem, which at low temperatures can be studied with the help
of the fixed-point Hamiltonian.\cite{AffleckLudwig93} Under these
conditions, the $s$-wave electrons, interacting via the isolated spin,
form the screening cloud. This many-body state still has the
Fermi-liquid properties, though its scattering characteristics are
different from those of just an isolated spin. One of the principal
differences is the shift of the scattering phase by $\pi/2$ for the
states at the Fermi level.\cite{Nozieres} This suggests another change
of the basis for the sake of convenience: from $s$-waves having the
scattering phase equal to $\pi/2$, $a_{k\sigma}^{(s)}$, to those with
the scattering phase $\pi$.

The formal relation between the $a_{k\sigma}^{(s)}$ and the new basis,
which we denote $b_{k\sigma}$, is given by
\begin{eqnarray}
b_{k\sigma}&\equiv&\int dx\, e^{ikx}
\hat{\Psi}_{\sigma}(x)
\;,\nonumber\\
\hat{\Psi}_{\sigma}(x)&=&\exp\left[i\pi \int_{-\infty}^x\!
  dx'\,g(x')\right] 
\hat{\psi}_{\sigma}^{(s)}(x)\;,\nonumber\\
\hat{\psi}_{\sigma}^{(s)}(x)&\equiv&\int\! dk\, e^{-ikx}
a_{k\sigma}^{(s)}\;,\label{fppsi}
\end{eqnarray}
where $g(x)$ is an arbitrary function obeying $\int_{-\infty}^\infty\!
dx'\,g(x')=1$.  The ``coordinate'' $x$ was introduced for convenience
to separate the incoming and outgoing parts of the scattering states,
which correspond to negative and positive values of $x$ respectively.

Before the scattering region [$x\to -\infty$ in Eq.~(\ref{fppsi})],
the wave functions of the states $b_{k\sigma}$ and $a_{k\sigma}^{(s)}$
coincide. Therefore the states
\begin{equation}
  \label{sptoinc}
  c_{k\sigma L}^{\rm (in)}=\xi b_{k\sigma}-\eta a_{k\sigma}^{(p)}\;,
\quad  c_{k\sigma R}^{\rm (in)}=\eta b_{k\sigma}+\xi a_{k\sigma}^{(p)}
\end{equation}
represent waves incident from a left or right lead only.  Passing the
scattering region, the wave function of the state $b_{k\sigma}$
acquires an extra phase of $\pi$ as compared to that of
$a_{k\sigma}^{(s)}$. Then one can see that the states
\begin{equation}
  \label{sptoout}
  c_{k\sigma L}^{\rm (out)}=-\xi b_{k\sigma}-\eta a_{k\sigma}^{(p)}\;,
\quad  c_{k\sigma R}^{\rm (out)}=-\eta b_{k\sigma}+\xi a_{k\sigma}^{(p)}
\end{equation}
have an outgoing wave only in the left or right lead respectively.
The current operator in terms of these states equals simply
\begin{equation}
  \label{fpI1}
\hat{I}(V)=\sum_{k,\sigma}\left(
c_{k\sigma L}^{{\rm (out)}\dagger}
c_{k\sigma L}^{{\rm (out)}\phantom{\dagger}}
-
c_{k\sigma R}^{{\rm (out)}\dagger}
c_{k\sigma R}^{{\rm (out)}\phantom{\dagger}}
  \right)\;.
\end{equation}

The fixed-point Hamiltonian in the basis $b_{k\sigma}$,
$a_{k\sigma}^{(s)}$ has a relatively simple
form:\cite{AffleckLudwig93}
\begin{eqnarray}
  \label{fpH}
\hat{H}_{\rm fp}&=&v_F\sum_{k\sigma} k
b_{k\sigma}^\dagger b_{k\sigma}
+v_F\sum_{k\sigma} k
a^{(p)\dagger}_{k\sigma}
a^{(p){\phantom{\dagger}}}_{k\sigma}
\nonumber\\
&-&\frac{v_F}{\nu T_K}\sum_{k_1k_2\sigma}(k_1+k_2)  
b_{k_1\sigma}^\dagger b_{k_2\sigma}^{\phantom{\dagger}}\nonumber\\
&+&\frac{1}{\nu^2 T_K}\sum_{k_1k_2k_3k_4}  
:b_{k_1\uparrow}^\dagger b_{k_2\uparrow}^{\phantom{\dagger}}
b_{k_3\downarrow}^\dagger b_{k_4\downarrow}^{\phantom{\dagger}}:
\;,
\end{eqnarray}
where $:...:$ denotes normal ordering. The spectrum of electrons
is linearized, $\varepsilon_k=v_F k$, since the reduced bandwidth is of
the order of $T_K\ll \varepsilon_F$; The Kondo temperature $T_K$ is
the only energy scale of the fixed-point Hamiltonian (\ref{fpH}).

The third term in Eq.~(\ref{fpH}) determines the phase shift which a
quasiparticle acquires as it passes through the dot. This shift is
energy-dependent: it equals $\pi$ at the Fermi level, as we discussed
before; and $\pi+\varepsilon_k/T_K$ in general case.\cite{Nozieres} In
terms of waves incident from the left or right lead, such behavior of
the phase shift is analogous to those in tunneling through a resonant
state tied to the Fermi level.  The fourth term in the Hamiltonian
describes the interaction of the quasiparticles of the Fermi liquid at
the fixed point. The $p$-waves are not affected by the Kondo
screening, so the Hamiltonian for them has the same form as the one
given by Eqs.~(\ref{H}), (\ref{LRtosp}).

Using the Hamiltonian~(\ref{fpH}), we can rewrite the current operator
(\ref{fpI1}) in the form more convenient for the further calculations:
\begin{equation}
  \label{fpI}
  \hat{I}(V)=(2\eta\xi)^2\left\{\frac{e^2}{\pi\hbar}V-\frac{ie}{\hbar}
\left[\hat{H}_{\rm fp},\sum_{k,\sigma}
c_{k\sigma R}^{{\rm (out)}\dagger}
c_{k\sigma R}^{{\rm (out)}\phantom{\dagger}}
\right]\right\}\;.
\end{equation}
The first term in Eq.~(\ref{fpI}) is the current that would flow if
all the states were able for resonant tunneling through the dot; the
scattering between the left- and right-incident species (which is just
backscattering when $\xi=\eta$) reduces the magnitude of the current,
and is accounted for by the second term.

To evaluate the conductance of the dot, we employ the Keldysh
technique (\ref{G}), treating the last two terms of Hamiltonian
(\ref{fpH}) as a perturbation.  At infinitely small temperature and
bias, the current through the dot is transferred by electrons at the
Fermi level. The transmission coefficient for these electrons equals
$(2\xi\eta)^2$, {\em i.e.} the second (backscattering) term in the
current operator (\ref{fpI}) yields zero.  Therefore the dot at this
conditions has maximum conductance, $G=G_U\equiv
(e^2/\pi\hbar)(2\eta\xi)^2$.  At finite temperatures, the electrons
which facilitate the current are spread within a strip of width $T$
near the Fermi level.  The departure of the particle energy from the
Fermi level in the system (\ref{fpH}) leads to the deviation of its
scattering phase from $\pi$, {\em i.e.}  from the resonance. Therefore
the conductance in this case will be lower than $G_U$.  Indeed,
substituting Eqs.~(\ref{fpI}) and (\ref{fpH}) into Eq.~(\ref{G}) and
employing the second order of the perturbation theory in the last two
terms of Hamiltonian~(\ref{fpH}), we arrive at
\begin{eqnarray}
  G_{\rm peak}(T)&=&\frac{e^2}{\pi\hbar}(2\eta\xi)^2
\left\{1-\frac{v_F}{\nu T_K^2}\sum_k k^2
  \left[-\frac{df(v_Fk)}{dk}\right]\right.\nonumber\\ 
&\quad&-\frac{2}{v_F\nu^3T_K^2}\sum_{k_1k_2k_3}
\left[-\frac{df(v_Fk_1)}{dk_1}\right]f(v_Fk_2)\nonumber\\
&\quad&\left.\left.\phantom{\frac{v_F^2}{\nu T_K^2}}
\times[1-f(v_Fk_3)]f[v_F(k_1-k_2+k_3)]\right]\right\}\nonumber\\
&=&
\left[1-\pi^2\left(\frac{T}{T_K}\right)^2\right]
G_U
\;,
  \label{fpGT}
\end{eqnarray}
where $f(\varepsilon)\equiv 1/[\exp(\varepsilon/T)+1]$ is the Fermi
distribution function. One can see from Eq.~(\ref{fpGT}) that the
conductance of the quantum dot system at low temperatures decreases
with growth of the temperature. This behavior has been observed
experimentally\cite{GoldhaberEtal98,CronenwettEtal98} and is analogous
to the decrease of the resistivity in a regular Kondo system (bulk metal
with magnetic impurities). 

The differential conductance of the dot at finite bias $V_{\rm
 dc}$, with $T\ll eV_{\rm dc}\ll T_K$, can be derived analogously to
Eq.~(\ref{fpGT}). The resulting formula
\begin{equation}
  \label{fpGV}
  G(V_{\rm dc})= 
\left[1-\frac38\left(\frac{eV_{\rm dc}}{T_K}\right)^2\right]
 G_U
\end{equation}
shows that $G(V_{\rm dc})$ decreases with the growth of the bias
applied to the dot.

Slow ($\hbar\omega\ll T_K$) ac field results in adiabatic
time-dependence of Kondo temperature, see Eq.~(\ref{TKt}).
The time-dependent part of the Hamiltonian (\ref{fpH}) 
with $1/T_K(t)$ given by Eq.~(\ref{TKt})
accounts for the interaction of quasiparticles with ac field.
To consider this part of Hamiltonian in the conventional terms of
electron-photon interaction, we expand $1/T_K(t)$ in Fourier series:
\begin{equation}
  \label{tTK}
  \frac{1}{T_K(t)}\equiv\sum_n\frac{1}{T_K^{(n)}}e^{in\omega t}
\end{equation}
After an act of photon absorption, a quasiparticle facilitating the current
is transferred from the Fermi level, {\em i.e.} away from the
resonance.  As a result, at low temperatures the ac field must
reduce the conductance of a quantum dot in the Kondo regime. At
$\hbar\omega\ll T_K$, the conductance can be calculated in the second
order of the perturbation theory in the time-dependent part of the
Hamiltonian. Substituting Eq.~(\ref{tTK}) into Eq.~(\ref{fpH}), and
then using the Keldysh formalism (\ref{G}) to evaluate the conductance
we arrive at
\begin{eqnarray}
G_{\rm peak}&=&
\left\{1-\sum_n
\left(\frac{1}{T_K^{(n)}}\right)^2
\left[
\frac{v_F^2}{\nu}\sum_k k^2
  \delta(v_F k-n\hbar\omega)
\right.\right.\nonumber\\
&&\quad+
\frac{2}{\nu^3}\sum_{k_1k_2k_3}
\delta(v_F k_1-n\hbar\omega)
\theta(v_Fk_2)\nonumber\\
&&\left.\left.\vphantom{\frac{v_F^2}{\nu T_K^2}}
\qquad\times[1-\theta(v_Fk_3)]\theta[v_F(k_1-k_2+k_3)]\right]\right\}
G_U
\nonumber\\
&=&
\left\{1-3\sum_n\left(\frac{\hbar n\omega}{T^{(n)}_K}\right)^2
\right\}G_U\;,
\label{Gnoznoz}
\end{eqnarray}
where for simplicity we set temperature to zero. Transforming
Eq.~(\ref{Gnoznoz}) back from $1/T_K^{(n)}$ to $1/T_K(t)$
[Eq.~(\ref{tTK})], we finally obtain
\begin{eqnarray}
G_{\rm peak}&=&
\left\{1-3\overline{
\left(\frac{d}{dt}
\frac{1}{T_K(t)}\right)^2}
\right\}G_U
\label{GNozfin}\\
&\approx&
\left\{1-\frac{3}{2}\left(\frac{\delta T_K}{T_K}\right)^2
\left(\frac{\hbar\omega}{T_K}\right)^2
\right\}G_U\;,
\nonumber
\end{eqnarray}
where $\overline{\vphantom{o}...\vphantom{o}}$ denotes averaging over
the period of variation of $T_K(t)$, and $\delta T_K/T_K$ is defined
by Eq.~(\ref{deltaTK}).

The single-photon decoherence processes described in Sec.~\ref{sec:decoh} do not
occur in this regime, because the energy necessary to flip the dot's
spin is increased by its interaction with the screening ``spin cloud''
in the leads, and is of the order of $T_K\gg \hbar\omega$. The rate
of the spin flip due to many-photon processes is exponentially small
in $T_K/\hbar\omega$.

\section{Scaling formula for the conductance}
\label{sec:sum1}

In this section we summarize the results obtained in
Secs.~\ref{sec:weak}--\ref{sec:strong} for the effect of the periodic
modulation of the dot's potential on the Kondo conductance.

In the absence of ac irradiation, the quantum dot system is described
by a number of physical parameters, see Eqs.~(\ref{HK}), (\ref{J0}).
However, in the Kondo regime all these parameters combine into a
single relevant energy scale, $T_K$, see Eq. (\ref{TK}), controlling
the behavior of the system, see, {\it e.g.}, Eqs.~(\ref{GRG}), and
(\ref{fpGT}). The periodic modulation $V_{\rm dot}\cos\omega t$ of the
dot potential adds two more parameters to the initial Hamiltonian
(\ref{HK}), and, most importantly, drives the system into a
non-equilibrium state. Surprisingly, such a drastic perturbation does
not break down the universal description of the problem, and the Kondo
temperature remains the only relevant energy scale.  We have shown
that the effect of the irradiation is described by two dimensionless
parameters $\hbar\omega/T_K$ and $\delta T_K/T_K\propto V_{\rm dot}$,
where $\delta T_K$ has the meaning of the adiabatic variation of the
Kondo temperature under the influence of ac modulation, see
Eq.~(\ref{deltaTK}).

At sufficiently large frequencies $\omega$ of the ac field, when 
\begin{equation}
  \label{lskdu}
\frac{\hbar\omega}{T_K}>\frac{32}{3\pi}
\frac{\left[\ln(\delta T_K/T_K)\right]^4}{\left[\delta T_K/T_K\right]^2}\;,
\end{equation}
the rate $\hbar/\tau$ of the spin-flip cotunneling exceeds the Kondo
temperature $T_K$.  The spin-flip cotunneling brings decoherence into
the spin dynamics of the dot, destroying the Kondo resonance. Small
lifetime of the Kondo resonance leads to a significant suppression of
the Kondo effect, see Sec.~\ref{sec:AffectedWeakKondo}. The dependence
of the zero-bias dc conductance $G_{\rm peak}$ of the dot on the power
and frequency of the ac field is given by Eqs.~(\ref{GRGtau}) and
(\ref{tauu}).

Upon lowering the frequency $\omega$, condition~(\ref{lskdu}) breaks
down, and $\hbar/\tau$ becomes smaller than the Kondo temperature.
Under such conditions, the strong suppression of the Kondo conductance
is not possible. However, the conductance still may deviate from the
unitary limit $G_U$. The violation of the condition~(\ref{lskdu})
occurs while $\hbar\omega$ still exceeds $T_K$.  The zero-bias
conductance in this regime can be estimated by
Eq.~(\ref{Gcross}) and (\ref{tauu}).

At frequencies below the Kondo temperature, the ac field is unable to
flip the spin of the dot, and the spin-flip cotunneling does not
occur. In this regime, the ac-driven deviation from the unitary limit
is small and can be accounted within the framework of the Fermi-liquid
description.\cite{Nozieres} The main role of the ac field is to scatter the
conduction electrons, transferring them to the energies off the Fermi
level. These scattered electrons miss the Kondo resonance, which is
tied to the Fermi level. It produces the small deviation of the dc
conductance $G_{\rm peak}$ from the unitary limit, see
Sec.~\ref{sec:strong}, Eq.~(\ref{GNozfin}).

\begin{figure}
\epsfxsize=8.5cm
\centerline{\epsfbox{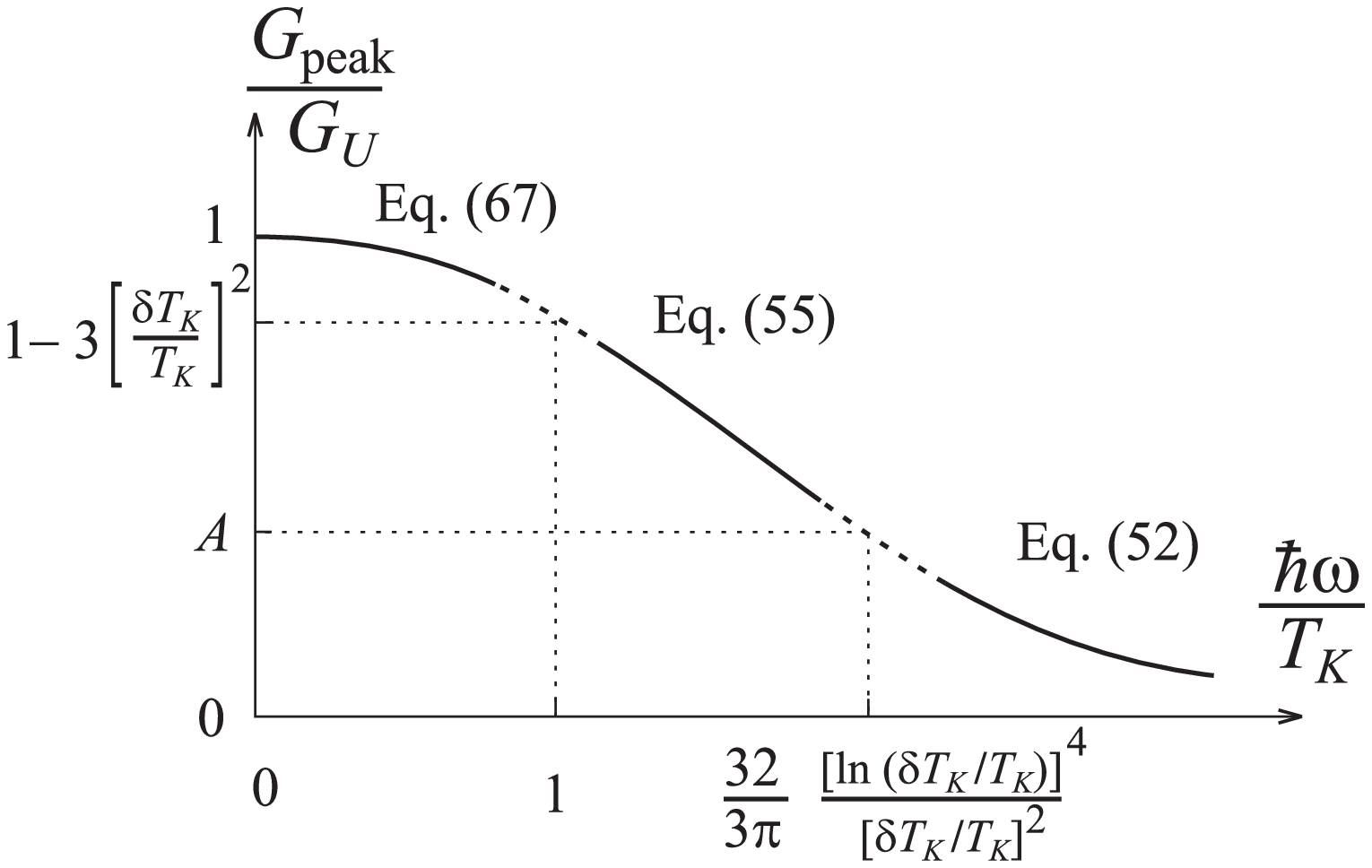}}
\caption{\label{fig:concl}
  The zero-bias Kondo conductance $G_{\rm peak}$ of a quantum dot
  monotonically decreases with the increase of the frequency
  $\hbar\omega$ of modulation of the dot potential. The plot shows the
  dependence of $G_{\rm peak}$ on $\hbar\omega$ at fixed amplitude
  $eV_{\rm dot}\propto \delta T_K/T_K$ of the modulation. The
  conductance in the unitary limit $G_U$ is given by
  Eq.~(\protect\ref{GU}).   The value of $G_{\rm peak}/G_U$ at
   $\hbar/\tau=T_K$ is denoted by $A\protect\lesssim 1$; this value can be
   found from numeric calculation.
}
\end{figure}

The results obtained for these three regimes match each other on the
corresponding limits of applicability. It allows us to piece together
the dependence of $G_{\rm peak}$ on $\delta T_K/T_K$ and $\hbar\omega$
in a broad frequency range, see Fig.~\ref{fig:concl}.  
\narrowtext

This dependence allows us to conjecture that at small $\delta
 T_K/T_K$ the conductance can be cast in the following form:
\begin{equation}
  \label{Guniv}
  \frac{G_{\rm peak}}{G_U}=F\left[\left(\frac{\delta T_K}{T_K}\right)^2
f\left(\frac{\hbar\omega}{T_K}\right)\right]\;,
\end{equation}
with two universal functions $F(x)$ and $f(y)$. Each of the functions
depends on only one variable; they have the following asymptotes
\begin{eqnarray}
  \label{F}
  F(x)=\left\{
    \begin{array}{ll}
1-ax\;,\quad & \mbox{if}\;x\ll1\;,\\
\displaystyle\frac{3\pi^2}{16}
\displaystyle\frac{1}{(\ln x)^2}\;,\quad & \mbox{if}\;x\gg1\;,
    \end{array}
\right.
\end{eqnarray}
and
\begin{eqnarray}
  \label{f}
  f(y)=\left\{
    \begin{array}{ll}
\displaystyle\frac{1}{a}3y^2\;\rule[-0.15in]{0in}{0.375in},\quad & \mbox{if}\;y\ll1\;,\\
\displaystyle\frac{3\pi}{32}
\displaystyle\frac{y}{(\ln y)^4}
\;,\quad & \mbox{if}\;y\gg1\;.
    \end{array}
\right.
\end{eqnarray}
The numerical parameter $a\sim 1$ is introduced and discussed in
Sec.~\ref{sec:crossover}. When $\hbar\omega\gg T_K$, the argument of
function $F$ has the meaning of dimensionless decoherence rate
$\hbar/\tau T_K$.

\section{Conductance suppression by ac bias}
\label{sec:weakacbias}

In sections \ref{sec:weak}--\ref{sec:sum1} we considered the effects
of modulation of the dot potential on the Kondo conductance.  In the
present section we study the conductance in the system where the
ac field is applied to the leads, thus creating alternating bias
$V_{\rm ac}$. The parameters characterizing such a field are the
dimensionless amplitude $eV_{\rm ac}/T_K$ and frequency $\hbar\omega'/T_K$.

First we consider the case of ``fast'' ac bias, 
$\hbar\omega'\gg\max\{T_K,\; eV_{\rm ac}\}$.
Under these conditions, the ac bias affects the Kondo conductance
through the decoherence of the dot's spin. The dependence of the
corresponding decoherence rate $\hbar/\tau'$ on the amplitude and
frequency of the ac bias can be calculated with the help of the
Renormalization Group technique which we used in
Sec.~\ref{sec:3rdorder}, \ref{sec:decoh}, and
\ref{sec:AffectedWeakKondo}.  The resulting expression reads
\begin{equation}
 \frac{\hbar}{\tau' T_K}=\frac{1}{\pi}
\frac{G_U}{e^2/\pi\hbar}
\left(\frac{eV_{\rm ac}}{T_K}\right)^2
\frac{T_K}{\hbar\omega'}
\frac{1}{[\ln(\hbar\omega'/T_K)]^2}\;.
\label{tauAClikedotRG}
\end{equation}
Note that, in contrast to the ac modulation of the gate voltage
(Sec.~\ref{sec:decoh}), in the case of ac bias the rate of decoherence
decreases with the growth of the field frequency $\omega'$.
The parameter
\begin{displaymath}
\frac{G_U}{e^2/\pi\hbar}
\equiv
\frac{4\Gamma_L\Gamma_R}{(\Gamma_L+\Gamma_R)^2}
\end{displaymath}
characterizes the asymmetry in the dot, and emerges in the expressions for
quantities associated with electron transfer between the leads.

When $\hbar/\tau'T_K>1$, the conductance can be evaluated by means of
the perturbation theory, see Sec.~\ref{sec:3rdorder}. The decaying
function $\langle S_j(t_1) S_k(t_2) S_l(t_3)\rangle$, which enters the
terms of the perturbation theory, provides the large-time cut-off for
the integrals in equations of the type (\ref{typterm}).  The
derivation of the expressions for the conductance is identical to the
one given in Sec.~\ref{sec:AffectedWeakKondo}, cf. Eq.~(\ref{GRGtau}).
The final formula reads
\begin{equation}
G_{\rm peak}=
\frac{3\pi^2}{16}
\frac{1}{\left[\ln(\hbar/\tau' T_K)\right]^2}
G_U\;,
\label{GRGtauprim}
\end{equation}
with $\hbar/\tau' T_K$ given by Eq.~(\ref{tauAClikedotRG}).

At smaller amplitudes, $\hbar/\tau'T_K<1$, the ac bias is
unable to suppress the formation of the Kondo many-electron state.
For this case we may repeat the reasoning of Sec.~\ref{sec:crossover}.
As the result, we obtain
\begin{equation}
  \label{Gcrossprim}
  G_{\rm peak}=
\left[1-a\frac{\hbar}{\tau' T_K}\right]G_U\;,
\end{equation}
{\em i.e.} the Kondo conductance is only weakly suppressed. 

In the opposite limit of slow variations of bias,
$\hbar\omega'\ll\max\{T_K,eV_{\rm ac}\}$, 
one can use the adiabatic approximation, 
\begin{equation}
  \label{Gacbiasadiabat}
  G_{\rm peak} = \overline{G(V_{\rm ac}\cos\omega't)}\;.
\end{equation}
Here $G(V)$ is the differential dc conductance at finite bias $V$,
and $\overline{\vphantom{o}...\vphantom{o}}$ denotes averaging over
the period of variation of ac bias.
For $eV/T_K\ll 1$, the conductance is given by
Eq.~(\ref{fpGV}). Substituting it  into
Eq.~(\ref{Gacbiasadiabat}), we obtain
\begin{equation}
G_{\rm peak}=
\left\{1-\frac{3}{16}\left(\frac{eV_{\rm ac}}{T_K}\right)^2
\right\}G_U\;.
\label{Gac1}
\end{equation}
In the opposite case, $eV_{\rm ac}/T_K\gg 1$, we obtain using
Eqs.~(\ref{GRGV}) and (\ref{Gacbiasadiabat}):
\begin{equation}
  \label{Gac2}
G_{\rm peak}=
\frac{3\pi^2}{16}
\frac{1}{[\ln (eV_{\rm ac}/T_K)]^2}G_U\;.
\end{equation}
  
\begin{figure}
\epsfxsize=8.5cm
\centerline{\epsfbox{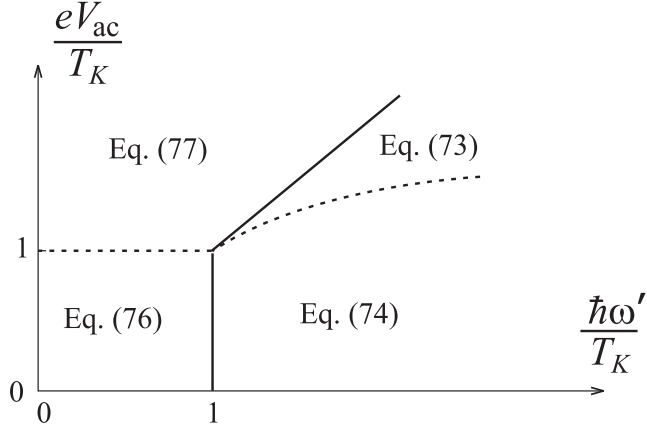}}
\caption{\label{fig:acmap}
  The regimes of the ac bias effect on the Kondo conductance.  The
  solid line is the border between the two domains
  (\protect\ref{region1}) and (\protect\ref{region2}). The dashed line
  separates the regimes of weak (below the line) and strong (above the
  line) suppression of the Kondo conductance in each of these two
  domains.
}
\end{figure}

Figure \ref{fig:acmap} shows the possible regimes of the ac bias
effect on the Kondo conductance of a dot. At
\begin{mathletters}
\begin{equation}
  \label{region1}
 \hbar\omega'\ll\max\{eV_{\rm ac},T_K\} \;,
\end{equation}
the peak conductance depends only on $eV_{\rm ac}/T_K$, see
Eqs.~(\ref{Gac1}),~(\ref{Gac2}). In the opposite case of high frequencies,
\begin{equation}
  \label{region2}
 \hbar\omega'\gg\max\{eV_{\rm ac},T_K\} \;,
\end{equation}
\end{mathletters}
the peak conductance depends only on $\hbar/\tau' T_K$, see
Eqs.~(\ref{GRGtauprim}), (\ref{Gcrossprim}).  Thus in both regions
(\ref{region1}) and (\ref{region2}), $G_{\rm peak}$ is a function of a
single variable. However, the corresponding variables are different in
the two regions. Therefore, at the crossover between these two
frequency domains, the peak conductance $G_{\rm
  peak}(\hbar\omega'/T_K, eV_{\rm ac}/T_K)$ can not be cast into a
simple form of a single-variable function.

\narrowtext

\begin{figure}
\epsfxsize=8.5cm
\centerline{\epsfbox{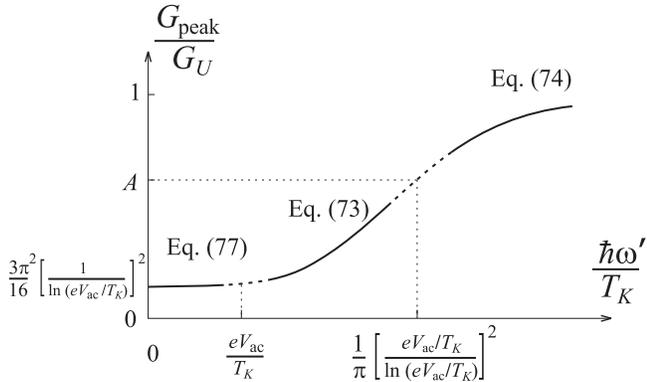}}
\caption{\label{fig:acconc}
The suppression of the Kondo effect with ac bias weakens, as the
frequency $\hbar\omega'$ of the bias grows. The plot shows 
the dependence of the dc zero-bias conductance $G_{\rm peak}$ on
$\hbar\omega'$ at fixed bias amplitude $V_{\rm ac}\gg T_K/e$. 
}
\end{figure}

It is instructive to consider the peak conductance as a function of
the frequency $\omega'$ at a fixed field amplitude $eV_{\rm ac}$. At
small amplitudes, $eV_{\rm ac}/T_K\ll 1$, the suppression of the Kondo
effect is weak at any frequency. At stronger fields, $eV_{\rm
  ac}/T_K\gg 1$, the Kondo effect is suppressed far below the unitary
limit at low frequencies.  The height of the zero-bias peak grows with
the increase of the field frequency $\omega'$, see
Fig.~\ref{fig:acconc}.  The description of the crossover between the regimes
(\ref{region1}) and (\ref{region2}) is developed in
Appendix~\ref{sec:acbiascross}.

\narrowtext

\section{Satellite conductance peaks}
\label{sec:sat}

In Sections \ref{sec:weak}--\ref{sec:weakacbias} of this paper we mostly
concentrated on the effects of the ac field on the zero-bias Kondo
conductance $G_{\rm peak}$. In this section we study how the ac field
modifies the finite-bias differential Kondo conductance $G(V_{\rm dc})$. 

Without an ac field, the dependence of the differential Kondo conductance
on $V_{\rm dc}$ is given by Eq.~(\ref{fpGV}) for $eV_{\rm dc}\ll T_K$
and by Eq.~(\ref{GRGV}) for $eV_{\rm dc}\gg T_K$. As we have seen from
the previous sections, the ac field reduces the height of the
zero-bias peak, $G_{\rm peak}\equiv G(V_{\rm dc}=0)$.

\begin{figure}
\epsfxsize=8.5cm
\centerline{\epsfbox{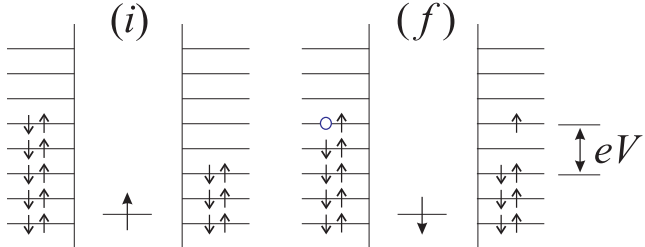}}
\caption{\label{fig:transDC}
 At finite bias, the coherence of the spin
state of the dot can be lost in an act of cotunneling, when an
electron leaves the dot to the lower voltage lead, and another
electron enters it from the higher voltage lead.
}
\end{figure}

Another effect of external irradiation on the differential conductance
$G(V_{\rm dc})$ is in producing satellite peaks at $eV_{\rm
  dc}=\pm n\hbar\omega$.  If an external ac field is applied, then, at
$eV_{\rm dc}=\pm n\hbar\omega$, a tunneling electron can hop from a state
at the Fermi level in one lead to a state at the Fermi level in the
other lead, emitting or absorbing $n$ photons. Thus at finite bias the
external irradiation can effectively put a tunneling electron into
zero-bias conditions, and the Kondo anomaly in the conductance is
revived. The height of these peaks can be calculated from the formula
(\ref{G}) similarly to Eq.~(\ref{Gt}). At low enough irradiation level,
$\gamma\ll 1$, it is sufficient to consider only one-photon
processes, accounted for by the Hamiltonian (\ref{HK}) with the
coupling constants ${\cal J}_{\alpha\alpha'}(t)$ given by
Eqs.~(\ref{Jreduced})--(\ref{gamma}). 
In this approximation, we will be able to describe the first pair of satellite
peaks, which emerge next to the main, zero-bias, peak.
 The resulting correction to
the conductance at $e|V_{\rm dc}|$ close to $\hbar\omega$ has the form
\begin{eqnarray}
&&G^{(3)}_{{\rm stl},\pm}(V_{\rm dc})=
\frac{3\pi^2}{8} \frac{e^2}{\pi\hbar} \nu^3\left[{\cal J}_{LR}^{\rm
(0)}\right]^2\left[{\cal J}_{RR}^{\rm (0)}+{\cal J}_{LL}^{\rm
(0)}\right]
\gamma^2 \nonumber\\&&\times
\int^0_{-\infty} \!\! dt  \,(-t) \exp (-|t|/\tau_{\rm stl})
\nonumber\\ &&\times
\left(\frac{\pi T}{\hbar}\right)^2
\frac{\cos [(eV_{\rm dc}\pm\hbar\omega)t/\hbar]}
{\sinh^2(\pi Tt/\hbar)+(T/D_0)^2}\;.
\label{Gtsat}
\end{eqnarray}
When $eV\ne\pm\,\hbar\omega$, the cosine function cuts off the
logarithmic divergence here. However, when $eV_{\rm
  dc}\to\pm\,\hbar\omega$, the cosine factor becomes essentially
constant [cf.  Eq.~(\ref{Gt}) at $V_{\rm dc}\to 0$], and the
differential conductance has a peak again. At $T\to 0$, the height of
the satellite conductance peak is determined by the spin decoherence
rate $\hbar/\tau_{\rm stl}$. We must mention that $\tau_{\rm stl}$ may
be significantly shorter than $\tau$ given by Eq.~(\ref{tau}). The
time $\tau$ characterizes the spin decoherence at zero bias, whereas
the satellite corresponds to a finite bias $eV_{\rm
  dc}=\pm\hbar\omega$. In the latter case, the spin decoherence occurs
mostly due to the tunneling of electrons through the dot (see
Fig.~\ref{fig:transDC}, and also Ref.~\onlinecite{MeirEtal93}). The
rate of this process at $\hbar\omega\gg T_K$ is given by
\begin{equation}
\frac{\hbar}{\tau_{\rm stl}}=\frac{1}{2\pi}\,\hbar\omega\,
\frac{\Gamma_L\Gamma_R}{\tilde{E}_d^2}\;.
\label{tausat}
\end{equation}
To extend the result (\ref{tausat}) to lower frequencies
$\hbar\omega\gtrsim T_K$, we employ the RG technique. As the result,
we obtain the decoherence rate as a function of the universal
parameter $\hbar\omega/T_K$:
\begin{equation}
\frac{\hbar}{\tau_{\rm stl}T_K}=\frac{3\pi}{32}\,\,
\frac{G_U}{e^2/\pi\hbar}
\frac{\hbar\omega/T_K}{[\ln(\hbar\omega/T_K)]^2}\;.
\label{tausatRG}
\end{equation}
Note that Eqs.~(\ref{tausat}) and (\ref{tausatRG}) imply $eV_{\rm
  dc}\approx \hbar\omega$.

\narrowtext

Using the RG technique, we derive the formula for $G_{{\rm
    stl},\pm}(V_{\rm dc})$, which is the contribution of one-photon
processes to the differential dc conductance:
\begin{eqnarray}
G_{{\rm stl},\pm}(V_{\rm dc})&=&
\left[
\ln\frac{\sqrt{(\hbar/\tau_{\rm stl})^2+
(eV_{\rm dc} \pm \hbar\omega)^2}}{T_K}+
\ln\frac{\hbar\omega}{T_K}
\right]^{-4}
\nonumber\\
&\times&
\frac{3\pi^2}{4}
\left[\frac{\delta T_K}{T_K}\right]^2
G_U\;.
\label{Gsat}
\end{eqnarray}
One can see from
Eqs.~(\ref{tausatRG})--(\ref{Gsat}) that $G_{\rm stl}$ depends on the
parameters of irradiation only through the universal variables $\delta
T_K/T_K$ and $\hbar\omega/T_K$. 
The details of derivation of Eq.~(\ref{Gsat}) are given in
Appendix~\ref{sec:satRG}. 

The full expression for $G(V_{\rm dc})$ reads
\begin{equation}
  \label{GVfull}
  G(V_{\rm dc})=G_{\rm main}(V_{\rm dc})
+G_{{\rm stl},+}(V_{\rm dc})+G_{{\rm stl},-}(V_{\rm dc})\;.
\end{equation}
Here $G_{\rm main}$ accounts for the tunneling through the dot without
absorption/emission of photons and is responsible for the zero-bias
Kondo peak.  The other two terms in Eq.~(\ref{GVfull}) describe the
satellite peaks in $G(V_{\rm dc})$.

As the criterion for resolution of the satellites, we adopt the
requirement that the function $G(V_{\rm dc})$ must be
non-monotonic on the sides of the zero-bias peak. This requirement
can be reformulated as:
\begin{equation}
  \label{satcond}
  \frac{\delta T_K}{T_K}>1\;,
\quad
\frac{\hbar\omega}{T_K}\gg 1\;.
\end{equation}
In the derivation of the condition (\ref{satcond}) we used
Eq.~(\ref{GVfull}) with $G_{\rm main}(V_{\rm dc})$ given by
Eq.~(\ref{GRGV}). Such a form of the elastic Kondo conductance $G_{\rm main}$
should be used because at $e|V_{\rm dc}|\sim\hbar\omega$ it is
suppressed mainly due to the finite bias rather than due to the
decoherence, since $\hbar\omega\gg \hbar/\tau_{\rm stl}\gg\hbar/\tau$.

The above consideration was performed for the ac field applied to the
gate. The case of ac bias is can be considered similarly. The
third-order perturbation theory result
for the shape of satellite peak may be obtained from
Eqs.~(\ref{Gtsat}) and (\ref{Gsat}) by replacing $\gamma^2$ with
$(\gamma')^2$. The RG treatment yields
\begin{eqnarray}
G_{{\rm stl},\pm}(V_{\rm dc})&=&
\left[
\ln\frac{\sqrt{(\hbar/\tau_{\rm stl})^2+
(eV_{\rm dc} \pm \hbar\omega)^2}}{T_K}+
\ln\frac{\hbar\omega}{T_K}
\right]^{-4}
\nonumber\\
&\times&\left[
\ln\frac{\hbar\omega}{T_K}
\right]^{4}
\frac{3\pi^2}{4}
\left[\frac{eV_{\rm ac}}{\hbar\omega'}\right]^2
G_U\;.
\label{Gsatacbias}
\end{eqnarray}
The condition for the satellite peaks to be clearly
visible takes the form
\begin{equation}
  \label{satcondacbias}
\frac{eV_{\rm ac}}{\hbar\omega'}>1\;,
\quad
\frac{\hbar\omega}{T_K}\gg 1\;.
\end{equation}

Conditions (\ref{satcond}) and (\ref{satcondacbias}), together with
Eq.~(\ref{GRGtau}), demonstrate that upon the increase of the
amplitude of the ac field, the zero-bias peak is suppressed first, and
only after that the satellite peaks may become distinguishable from
the background conductance.

\section{Comparison with experiment}
\label{sec:exp}

In this section, we discuss possible ways of comparison of the theory
presented here with experiments. Ideally, one should measure the dc
conductance through a dot while applying the ac bias to the gates or
leads in a controllable way, and varying its frequency $\omega$ and
amplitude in a broad range. Our results predict that the data obtained
at various values of $T_K$ should be scalable, when using the proper
dimensionless variables, Eqs.~(\ref{F}), (\ref{f}),
(\ref{GRGtauprim}), (\ref{Gcrossprim}), (\ref{Gac1}), and
(\ref{Gac2}). We predict also that at a fixed magnitude of ac field,
the suppression of the Kondo effect becomes more severe with the
increase of $\omega$, if the applied field modulates the gate
potential; the dependence on $\omega$ in the case of ac field applied
to the leads is opposite. If $\omega$ significantly exceeds $T_K$,
see Eqs.~(\ref{satcond}) and (\ref{satcondacbias}), observation of
``satellites'' at $eV=\hbar\omega$ of the main Kondo singularity may
become possible.  The appearance of even small side peaks though is
associated with a strong suppression of the zero-bias Kondo
singularity.

Presently, there is only one experiment aimed at observation of
effects of irradiation on the Kondo conductance in the quantum
dot.~\cite{exp} In the analysis of this experiment below, we will see
that the frequencies used were of the order of $T_K$, and the range of
$\omega$ was less than one decade. In addition, it was impossible to
calibrate the amplitude of the field applied to the device; the
attenuation coefficients were frequency-dependent. Therefore, one
could not perform the measurements of $G_{\rm peak}(\omega)$ at a
fixed field amplitude.  However, it was possible to measure the
dependence of the peak conductance on the amplitude of the applied
field at a discrete set of fixed frequencies. The ac source was
powerful enough to allow the authors of Ref.~\onlinecite{exp} to
completely suppress the Kondo anomaly. Nominally, the ac field was
applied to the gate, but it was apparently hard to exclude ``leaking''
of the field to other electrodes of the
device.~\cite{GoldhaberPrivcom} It creates further ambiguity in the
interpretation of the experimental results.

The Kondo temperature of the system can be found from the dependencies
$G_{\rm peak}(T)$ and $G(V_{\rm dc})$ measured in the absence of
irradiation. Comparing the experimental results with the theory
[Eqs.~(\ref{fpGT}) and (\ref{fpGV})], we obtain the estimate $T_K\sim
30-60$ $\mu$eV.  The external irradiation was applied through a
high-frequency coaxial cable, coupled capacitively to the gate. The
frequency of the irradiation ranged from 10 GHz to 50 GHz ({\em i.e.}
$\hbar\omega$ between 40$\mu$eV and 200 $\mu$eV).

The zero-bias Kondo peak was clearly observed in the $G(V_{\rm dc})$
dependence, when no ac field was applied.  With the increase of the
amplitude of the ac field, the height of the zero-bias peak decreased
for each of the frequencies used. Such a behavior is in agreement with
the our conclusion that irradiation must suppress the Kondo effect in
a quantum dot even when dot ionization does not occur. Also, the
satellites did not appear, in agreement with our theory for
$\hbar\omega\sim T_K$, see Sec.~\ref{sec:sat}.

The authors of Ref.~\onlinecite{exp} attempted the data collapse for
the dependence of $G_{\rm peak}$ on the amplitude $V_{\rm irr}$ and
frequency $\omega$ of the ac field. In this procedure, $V_{\rm irr}$
was scaled by some frequency-dependent parameter, in order to bring to
a single curve the dependencies $G_{\rm peak}$ {\it vs.} $V_{\rm irr}$
measured at different values of $\omega$. Successful data collapse
means that a universal dependence exists,
\begin{equation}
  \label{exper}
  G_{\rm peak}=F[V_{\rm irr}f(\hbar\omega)]\;.
\end{equation}
In the experiment~\cite{exp} the scaled curves coincided with good
precision, see Fig.~7 of Ref.~\onlinecite{exp} for conductances
$G_{\rm peak}$ ranging from the unitary limit down to values small compared
with $G_U$. Our theory indeed allows for such a
single-parameter scaling~(\ref{exper}), see Eqs.~(\ref{F}) and
(\ref{f}) and Fig.~\ref{fig:concl}, in the case of ac modulation of
the gate voltage. On the contrary, according to Fig.~\ref{fig:acmap}
scaling of the type (\ref{exper}) would be possible only in the domain
of large or small conductances, if the ac field of frequency
$\omega\sim T_K$ is applied to the leads.

We must mention here that the experimentally measured\cite{exp} Kondo
conductance was suppressed by the irradiation uniformly across the
Coulomb blockade valley, including its middle point, where
$E_d=U-E_d$.  In our theory, however, the suppression is not uniform,
and vanishes at $E_d=U/2$, if the ac field is applied to the gate
[see Eqs.~(\ref{gamma}),~(\ref{tauu}), and (\ref{deltaTK})].  We see
two possible reasons for the discrepancy.  First, our conclusion is
valid for a weak modulation only. This condition most probably was not
satisfied in the experiment\cite{exp} performed at $\hbar\omega\sim T_K$; at
such frequencies  strong modulation of the gate potential,
$\delta T_K\sim T_K$, is required to achieve  significant suppression
of the Kondo effect. Second,  leakage of the ac field to the leads
would result in  suppression of the Kondo conductance by ac
bias. This effect is not so sensitive to a specific value of $E_d$,
see Section~\ref{sec:weakacbias}.

\section{Conclusion}

We considered the Kondo conductance of a quantum dot subjected to ac
field. We have shown that, despite the essentially non-equilibrium
character of the problem, the Kondo temperature $T_K$ [Eq.~(\ref{TK})]
remains the only relevant energy scale. The dc Kondo conductance
depends on the ac field only through two dimensionless
variables which are the frequency and the amplitude of the ac
perturbation divided by $T_K$. In terms of these two variables,
conductance is a universal function. The form of this function, and
the relation of the perturbation amplitude to the ``bare'' value of
the ac field amplitude, depends on the way the ac field is applied.

If the ac field is applied to the gate, then the strength of the
perturbation is characterized by the amplitude $\delta T_K$ of
adiabatic variations of the Kondo temperature, see
Eqs.~(\ref{deltaTK})--(\ref{tildeEdt}). At low frequencies,
$\hbar\omega<T_K$, the conductance is close to the unitary limit
[Sec.~\ref{sec:strong}, Eq.~(\ref{GNozfin})]. At higher frequencies,
$\hbar\omega>T_K$, the ac field suppresses the Kondo effect by means
of the decoherence of the dot's spin (Sec.~\ref{sec:weak},
\ref{sec:crossover}). The value of the zero-bias conductance decreases
with the increase of the frequency $\omega$ of the ac field. The
results we obtained for the modulation of the gate voltage are
summarized in Sec.~\ref{sec:sum1}.

If the ac field is applied across the dot, then the proper variable
is the corresponding dimensionless bias between the leads, $V_{\rm
  ac}/T_K$. A ``slow'' field, $\hbar\omega'<\max\{eV_{\rm ac},
T_K\}$, suppresses the Kondo effect essentially the same way as a
finite dc bias does, see Eqs.~(\ref{Gac1}), (\ref{Gac2}).  A ``fast''
ac field, $\hbar\omega'>\max\{eV_{\rm ac}, T_K\}$, affects
the Kondo conductance through the decoherence of the dot spin
[Eqs.~(\ref{tauAClikedotRG}), (\ref{GRGtauprim}), (\ref{Gcrossprim})].
At the fixed amplitude of the field, the suppression of the Kondo
effect diminishes with the increase of the ac field frequency
$\omega'$.

The ac field also produces satellite peaks in the dependence of
the differential dc conductance on the dc bias. However, the satellite
maxima in the conductance are inevitably small, see Sec.~\ref{sec:sat}.

The analysis of the experiment\cite{exp}
[Sec.~\ref{sec:exp}] demonstrates good agreement between our
theoretical results and the results of the experiment.

\acknowledgments

The work at the University of Minnesota was supported by NSF Grant DMR
97-31756. LG acknowledges the hospitality of the Delft University of
Technology. LG and AK acknowledge also the hospitality of Institute of
Theoretical Physics supported by NSF Grant PHY 94-07194 at University
of California at Santa Barbara, where a part of the work was
performed. The authors are grateful to L.P.~Kouwenhoven,
D.~Goldhaber-Gordon and Y.~Meir for useful discussions.

\appendix

\section{ac bias with $\protect\lowercase{e}V_{\rm \protect\lowercase{ac}}\sim \hbar\omega'$}
\label{sec:acbiascross}

In this Appendix we describe the crossover between the regimes of
``slow'' and ``fast'' ac bias, 
which occurs at 
\begin{equation}
  \label{appBcond}
\hbar\omega'\sim eV_{\rm ac}\gg T_K\;,
\end{equation}
see Fig.~\ref{fig:acmap}. Throughout this Appendix, we will use the
finite-order perturbation theory to evaluate the decoherence rate and
conductance. The RG technique is abandoned here, since the
finite-order perturbation theory is sufficient in the region defined
by condition (\ref{appBcond}).

\subsection{Decoherence by ac bias}
\label{sec:actau}

Unlike the case $\gamma'\equiv eV_{\rm ac}/\hbar\omega'\ll 1$,
in the crossover region (\ref{appBcond}) the decoherence rate is
determined also by many-photon processes.
Using the Fermi Golden Rule with the
Hamiltonian of Eqs.~(\ref{HK}) and (\ref{acbiasJfull}), we arrive at
\begin{equation}
\qquad\frac{\hbar}{\tau'}=\frac{2}{\pi}
\frac{\Gamma_L\Gamma_R}{\tilde{E}_d^2}
\,M\!\left(\gamma'\right)
\hbar\omega'\;,\label{tauAC}\\
\end{equation}
where $M(x)\equiv -xJ_0(x)J_1(x)
+x^2\left[J_0(x)\right]^2
+x^2\left[J_1(x)\right]^2$.
In the case of ``fast'' ac field, $\gamma'\equiv eV_{\rm
  ac}/\hbar\omega'\ll 1$, equation (\ref{tauAC}) reduces to
\begin{equation}
 \frac{\hbar}{\tau'}=\frac{1}{\pi}
\frac{\Gamma_L\Gamma_R}{\tilde{E}_d^2}
\left(\gamma'\right)^2
\hbar\omega'\;.
\label{tauAClikedot}
\end{equation}
The latter formula is similar to Eq.~(\ref{tau}) and accounts for
single-photon processes only. Equation (\ref{tauAClikedot}) is the
first term in the perturbation theory series in ${\cal
  J}_{\alpha\alpha'}$ for $\hbar/\tau'$. The summation of leading terms of
all orders in ${\cal J}_{\alpha\alpha'}$ can be performed with the
help of the RG technique, and yields Eq.~(\ref{tauAClikedotRG})
[cf. Eqs.~(\ref{tau}) and (\ref{tauu}) respectively].

In the limiting case of ``slow'' ac bias, $\gamma'\gg 1$,
Eq.~(\ref{tauAC}) is reduced to
\begin{equation}
\frac{\hbar}{\tau'}=\frac{4}{\pi^2}
eV_{\rm ac}
\frac{\Gamma_L\Gamma_R}{\tilde{E}_d^2}
\;,
\label{tauAClikeDC}
\end{equation}
cf. Eq.~(\ref{tausat}). 

\subsection{Conductance}

The conductance in the crossover region (\ref{appBcond})
can be evaluated with the
third-order perturbation-theory series in ${\cal
  J}_{\alpha\alpha'}(t)$.  Using Eqs.~(\ref{HK}), (\ref{acbiasJfull}),
(\ref{G}), and (\ref{I}), we arrive at
\begin{eqnarray}
&&G_{\rm peak}^{(3)}(T,V_{\rm dc})=\frac{3\pi^2}{2} \frac{e^2}{\pi\hbar}
\nu^3\left[{\cal J}_{LR}^{\rm (0)}\right]^2
\left[{\cal J}_{RR}^{\rm (0)}+{\cal J}_{LL}^{\rm (0)}\right]
\nonumber\\
&&\quad
\times\int^0_{-\infty}\!\! dt\,
\frac{(-t)\cos (eV_{\rm dc}t/\hbar)\exp(-|t|/\tau)}
{\sinh^2(\pi Tt/\hbar)+(T/D_0)^2}\left(\frac{\pi T}{\hbar}\right)^2
\label{Gtac}\\
&&\qquad\times\cos\left[
\gamma'\sin(\omega't+\phi_0)-
\gamma'\sin\phi_0\right]\nonumber
\end{eqnarray}
[cf. Eq.~(\ref{Gt})].  The expression
(\ref{Gtac}) accounts for the current induced by both dc and ac
biases. To single out the former contribution, which is the true dc
conductance, we must average over the phase $\phi_0$. This averaging
in the limit of zero temperature and dc bias yields
\begin{eqnarray}
G_{\rm peak}^{(3)}&=&\frac{3\pi^2}{2} \frac{e^2}{\pi\hbar}
\nu^3\left[{\cal J}_{LR}^{\rm (0)}\right]^2
\left[{\cal J}_{RR}^{\rm (0)}+{\cal J}_{LL}^{\rm (0)}\right]
\nonumber\\
&&\quad
\times\int^0_{-\infty}\!\! dt\,
\frac{\exp(-|t|/\tau)}{\sqrt{t^2+(\hbar/\pi D_0)^2}}
J_0\left[2\gamma'\sin\frac{\omega't}{2}\right]\;.
\label{G3acbiasfinal}
\end{eqnarray}

In the limit $\omega'\to 0$, equation (\ref{G3acbiasfinal})
yields
\begin{equation}
  \label{eqqq}
  G_{\rm peak}^{(3)}\propto\ln(D_0/eV_{\rm ac})\;.
\end{equation}
This result is analogous to Eq.~(\ref{Gac2}) at $eV_{\rm ac}\gg T_K$.
At small frequencies, $\omega'\tau<1$, the corrections to
Eq.~(\ref{eqqq}) are proportional to $\exp(-1/\omega'\tau)$.  At
larger frequencies $\hbar/\tau<\hbar\omega'\ll eV_{\rm ac}$, the two
leading terms in the expansion of the right-hand side of
Eq.~(\ref{Gtac}) in powers of $1/\gamma'$ are
\begin{eqnarray}
  \label{G3acbiasVac}
G_{\rm peak}^{(3)}&=&\frac{3\pi^2}{32} \frac{e^2}{\pi\hbar}
\nu^3\left[{\cal J}_{LR}^{\rm (0)}\right]^2
\left[{\cal J}_{RR}^{\rm (0)}+{\cal J}_{LL}^{\rm
    (0)}\right]\nonumber\\
&&\times
\left[
\ln\frac{D_0}{eV_{\rm ac}}
+\frac{1}{\gamma'}\ln\omega'\tau
\right]\;.
\end{eqnarray}

At even larger frequencies, $\hbar\omega'\gg eV_{\rm ac}$,
Eq.~(\ref{G3acbiasfinal}) yields 
\begin{equation}
  \label{eqww}
 G_{\rm peak}^{(3)}\propto\ln(D_0\tau'/\hbar) \;.
\end{equation}
This is the first logarithmic term of the series that are summed up in
Eq.~(\ref{GRGtauprim}). 

Note that the results given by Eqs.~(\ref{eqqq}), (\ref{G3acbiasVac}),
and (\ref{eqww}) match each other at the corresponding applicability
limits.

\section{RG transformation for satellite peaks}
\label{sec:satRG}

In this Appendix, we describe the RG transformation we used to derive
Eq.~(\ref{Gsat}).  Unlike the other instances of application of the RG
technique in our paper, the RG transformation of this Appendix
consists of two stages.  The first stage is analogous to the one
considered in Sec.~\ref{sec:decoh} [Eqs.~(\ref{tauRGJ}),
(\ref{RGsolutprim})].  It stops when the bandwidth $D$ reaches
$eV_{\rm dc}\approx\hbar\omega$. Since $\hbar/\tau_{\rm
  stl}<\hbar\omega$ [see Eq.~(\ref{tausatRG})], we need to reduce the
band further to account for all possible virtual transitions
contributing to the Kondo anomaly in $G_{\rm stl}(V_{\rm dc})$.

The RG transformation we consider in this Appendix is aimed at
evaluation of the conductance $G_{\rm stl}$. In processes which contribute
to the singularity in $G_{\rm stl}$, a tunneling electron jumps from
the Fermi level in one lead to the Fermi level in the other, emitting
or absorbing a photon. Therefore, the reduced band in each lead must
be centered at its Fermi level.  When $D$ is below
$\hbar\omega$, transitions of only two types are possible within such
a band. First, there can be transitions within a lead without
absorption/emission of a photon. Second, transitions from the
higher-potential lead to the lower-potential lead with emission of a
photon, and reverse transition with absorption of a photon are also
possible.  The other types of transitions bring electrons out of the
reduced band and should be excluded from the consideration. Such a
treatment yields the RG equations
\begin{eqnarray}
  \label{RGeqsat}
  \frac{d{\cal J}_{\alpha\alpha}}{dD}&=&\nu\frac{{\cal
  J}_{\alpha\alpha}^2}{D}\;,\\
  \frac{d{\cal J}_{LR}}{dD}&=&
\nu\frac{{\cal J}_{LR}({\cal J}_{LL}+{\cal J}_{RR})}{D}\;,
\end{eqnarray}
with the initial conditions
\begin{eqnarray}
\left.{\cal J}_{\alpha\alpha}(D)\right|_{D\sim\hbar\omega}&=&
\frac{2\Gamma_\alpha}{\Gamma_L+\Gamma_R}
    \frac{1}{2\nu\ln(\hbar\omega/T_K)}
\;.\\
\left.{\cal J}_{LR}(D)\right|_{D\sim\hbar\omega}&=&
\frac{2\sqrt{\Gamma_L\Gamma_R}}{\Gamma_L+\Gamma_R}
    \frac{1}{4\nu[\ln(\hbar\omega/T_K)]^2}
\frac{\delta T_K}{T_K}  
\;.
  \label{RGinitsat}
\end{eqnarray}
cf. Eq.~(\ref{RGsolutprim}). The second stage of the transformation
must be stopped at $D\sim D^*\equiv \sqrt{(\hbar/\tau_{\rm
    stl})^2+(eV_{\rm dc}\pm\hbar\omega)^2}$. Expanding the solution
for ${\cal J}_{LR}$ in powers of $\delta T_K/T_K$ up to the first
power, we obtain
\begin{eqnarray}
&&\left.{\cal J}_{LR}\right|_{D\sim D^*}=
\frac{2\sqrt{\Gamma_L\Gamma_R}}{\Gamma_L+\Gamma_R}
\frac{\delta T_K}{T_K}\nonumber\\
&&\times\left[
\ln\frac{\sqrt{(\hbar/\tau_{\rm stl})^2+
(eV_{\rm dc} \pm \hbar\omega)^2}}{T_K}+
\ln\frac{\hbar\omega}{T_K}
\right]^{-2}
\;.
  \label{sfgfgh}
\end{eqnarray}
The conductance $G_{\rm stl}$ must be calculated in the second-order
perturbation theory in ${\cal J}_{LR}(D\sim D^*)$, given by
Eq.~(\ref{sfgfgh}). This calculation finally yields Eq.~(\ref{Gsat}).

\end{multicols}
\end{document}